\documentclass[aip,pop,reprint,amsmath,amssymb,nofootinbib]{revtex4-1}
\pdfoutput=1
\usepackage{geometry}
\usepackage{color}
\usepackage{calc}
\usepackage{amsmath}
\usepackage{amssymb}
\usepackage{cancel}
\usepackage{graphicx} 
\usepackage{esint}
\PassOptionsToPackage{normalem}{ulem}
\usepackage{ulem}
\usepackage{sidecap}
\makeatletter

\usepackage{bm}
\usepackage{soul}
\usepackage{url}
\usepackage{varioref}
\usepackage{graphics}
\usepackage{subfigure}
\usepackage{adjustbox}
\usepackage{appendix}
\usepackage{multirow}
\usepackage{verbatim}
\usepackage{colortbl}
\usepackage{times}
\usepackage{multimedia}
\usepackage{enumerate}
\usepackage{dcolumn}
\usepackage[usenames,dvipsnames,svgnames,table]{xcolor}
\usepackage{wrapfig}
\usepackage{nth}

\newcommand{\mf}{\mathbf}
\newcommand{\be}{\begin{equation}}
\newcommand{\ba}{\begin{eqnarrray}}
\newcommand{\ee}{\end{equation}}
\newcommand{\ea}{\end{eqnarray}}

\def\th{\theta}

\def\part{\partial}

\newcommand{\bmin}{\begin{minipage}{0.495\textwidth}}
\newcommand{\emin}{\end{minipage}}

\def\Talf{\tau_{A}}

\newcommand{\vvec}{\mathbf{v}}
\newcommand{\Bvec}{\mathbf{B}}
\newcommand{\Evec}{\mathbf{E}}
\newcommand{\jvec}{\mathbf{j}}

\def\vperp{\vvec_{\perp}}

\begin{document}

\title{Properties of the electrostatically driven helical plasma state}
\author{Cihan Ak\c{c}ay}
\email{c\_akcay@tibbartech.com}
\author{John M.~Finn, Richard A.~Nebel, Daniel C.~Barnes, and Neal Martin}
\affiliation{Tibbar Plasma Technologies, 274 DP Rd., Los Alamos, NM 87544}

\begin{abstract}
A novel plasma state has been found [C.~Ak\c{c}ay, J.~Finn, R.~Nebel and D.~Barnes, 
Phys.~Plasmas \textbf{24}, 052503 (2017)] in the presence of a uniform applied
axial magnetic field in periodic cylindrical geometry. This state
is driven by external electrostatic fields provided by helical electrodes, and depends on radius $r<r_w$ and $m\theta-n\zeta$, 
where $m=n=1$, $\theta$ is the poloidal angle, and $\zeta=z/R$ is the toroidal angle. 
In this reference, the strongly driven form of the state was found to have a strong axial mean 
current density, with a mean-field line safety
factor $q_0(r)$ just above the pitch of the electrodes $m/n=1$ in the interior, 
where
the plasma is nearly force-free. However, at the edge the current density has a component 
perpendicular to $\mathbf{B}$. This perpendicular current density drives nearly 
Alfv\'enic 
helical plasma flows, an notable feature of these states. This 
state is being studied for its possible application to DC electrical transformers 
and possibly tailoring the 
current profile in tokamaks. We present results on several issues of importance for these applications: 
the transient leading to the steady state; the twist and writhe of the field lines and 
their relation with the current density; the properties of the current 
density streamlines and length of the current density lines connected to the electrodes; 
the sensitivity to changes in the velocity boundary conditions; 
the effect of varying the radial resistivity profile; and the effects of a concentrated electrode 
potential.

\end{abstract}
\maketitle

\section{Introduction\label{sec:Introduction}}

In a previous paper\cite{Akcay2017A}, hereafter referred to as AFNB, the zero-pressure resistive MHD framework 
was employed to describe the physics of a periodic cylindrical plasma that is 
driven electrostatically by helical electrodes located at the radial boundary. 
The drive at the wall $(r=r_w)$ was specified by the electrostatic potential as $\phi_{0}e^{im\theta+ikz}+c.c.=
\phi_{0}e^{i\theta-i\zeta}+c.c.$, and
the normal magnetic field at the wall, $B_r(r_w)$ was taken to be zero. Here, $k=-n/R$, $\zeta$ is the toroidal angle, $\zeta=z/R$, $R$ is the major radius 
(periodicity length $L=2 \pi R$), and $m=n=1$.
For a small electrostatic drive $\phi_{0}$, the time-asymptotic state was found to consist of the initial uniform field $B_z$ plus 
a linear $m=n=1$ perturbation. 
The mean poloidal magnetic field was observed to be very small, yielding a total current $I_z\approx 0$. 
For a larger drive $\phi_0$, the time-asymptotic state was observed to be a single helicity Ohmic steady state, with a broad 
spectrum of $(m,n)$ but still with $m/n=1$, similar to the quasi-single 
helicity (QSH) states\cite{FinnNebelBathke,EscandeQSH} and specifically single helical axis  (SHAx) states\cite{SHAX-1,SHAX-2} 
(without magnetic islands) in reversed field pinches.
This state was observed to have highly distorted flux surfaces (surfaces of constant helical flux $\chi=mA_z-krA_{\theta}$) 
with nearly Alfv\'enic helical flows and a flat 
quasilinear safety factor profile 
$q_{0}=rB_{\theta}^{(0,0)}/RB_{z}^{(0,0)}\gtrsim m/n=1$ except near the plasma 
edge. 
Another important characteristic of this more strongly driven state is that both the flux surface average $\langle 
\eta\mathbf{j\cdot B}\rangle=\langle \eta\lambda B^{2}\rangle$ and $\lambda$ 
on the magnetic axis ($O$-line) are zero, as discussed in AFNB.
The former implies that $\lambda=\mathbf{j\cdot B}/B^2$ consists only of Pfirsch-Schl\"uter currents\cite{Pfirsch1962}, 
and for $\beta=0$ these are related to inertial and viscous stresses perpendicular to the magnetic field rather than pressure 
gradients.
The flux surface average condition was also shown in AFNB to be consistent with a constant 
magnetic helicity\cite{MyRelativeHelicity} $K_p$ in the time-asymptotic state. 
AFNB concluded that while there is no helicity injection from the boundary ($\dot{K}_{inj}=0$), because $B_r(r_w)=0$, the helical plasma self-generates magnetic helicity during the transient stage in the region where $\lambda<0$ \textit{via} the term traditionally associated with only the resistive dissipation of helicity. 
For a larger potential, $\phi_{0}>\phi_{crit}$, the time-asymptotic state is no 
longer steady; for practical purposes the operating range of helical 
potential $\phi_0$ is the interval between the value for which $q_0(r=0)\gtrsim 1.0$ and $\phi_{crit}$, where, according to AFNB the perpendicular velocity is comparable
to the Alfv\'{e}n speed. 
AFNB also showed that the aforementioned characteristics arise in simulations where the 
helical drive is applied as a normal current density source $j_r$ at the wall instead of a potential $\phi_0$. 

AFNB also found some of the properties of the above time-asymptotic state to be insensitive to the resistivity 
profile and velocity boundary condition while other properties exhibited a strong sensitivity, most notably 
the surfaces of the helical field $g=mB_z -krB_{\theta}$, the analog of the helical flux.
It is the sensitivity of $g$ and in general the current density streamlines 
that is the focus of Sec~\ref{sec:Sensitivity} of this paper. 
The application of a loop voltage (back EMF), $E_{0}L$ to 
simulate the effect of a secondary circuit was also investigated in AFNB, as was the dependence on the Lundquist number $S$.

Two possible applications of this unique plasma steady-state were described in AFNB. 
These are (1) the development of direct current (DC) electrical transformers\cite{Nebel2015,Nebel2016} and (2) 
the possibility of tailoring the current density profile in a tokamak or a reversed field
pinch (RFP). 
We focus mainly on the first application in this publication. 

In this paper we focus on further investigations of the properties of the 
helical plasma state, starting first with the transient stage that 
leads to the final time-asymptotic state of AFNB in the strong drive regime.
The results indicate that the early transient stages have approximately equal distributions of positive and negative 
$\lambda$ ($\approx j_z$), as 
expected when the perturbation is small enough to be in the linear regime. 
These results also show a very small increase in the magnetic helicity contained in the 
volume $K_p$, also consistent with the approximate linearity and $\dot{K}_{inj}=0$.
Later in the transient, as the perturbation becomes stronger, the rate of change of magnetic helicity $\dot{K}_p$ grows, 
with positive contributions in the regions where $\lambda<0$ ($\mathbf{j}\cdot\mathbf{B}<0$) and losses where $\lambda>0$.
As argued in AFNB,  $\dot{K}_p$ is $-2$ times the flux surface average $\langle \eta \mathbf{j\cdot B} \rangle=\langle 
\eta \lambda B^2\rangle$, 
integrated over a flux surface variable, so the flux surface
average condition for the time-asymptotic state in AFNB is violated during this transient period, 
as it is in the presence of back EMF, as discussed in AFNB. 
The energy dissipation channels during the transient are also presented here for the nominal state of AFNB. 
Our findings indicate that the input power is dissipated mainly Ohmically during the transient, and by viscous mechanisms in the time-asymptotic state. 
However, this behavior is sensitive to the velocity boundary condition at the radial wall. 
For example, viscous dissipation is negligible for cases that impose a homogeneous Neumann velocity boundary condition 
because of the resulting (nearly) flat velocity profiles. 
The locations of the helical $O-$point $r_0$, the mean field rotational transform
$1/q_0(0)$, and the bulk velocity $||\vperp||/v_A$are all tracked as a function of time during the transient.

The twist and writhe of the field lines on the helical flux surfaces and their relationship with the current density are 
also investigated.
We quantify how the twist and writhe characterize the magnetic field lines on the helical flux surfaces 
in a more representative way 
than the quasilinear $q_0(r)$.

A major topic of this paper is the study of the properties of the steady-states obtained in terms of two metrics related to the current density streamlines. The first 
concerns the existence of current that directly flows between secondary electrodes at the ends; the second involves
the current that leaks or ``shunts'' between the primary electrodes at $r=r_w$ and the secondary electrodes.
Metric (I), described in Sec.~\ref{sec:g-field-terms}, focuses on the helical field $g$ and the terms from the axial component of Ohm's law that contribute to the evolution of $g$ in steady state.
The current density streamlines lie on surfaces of constant $g$, as discussed in AFNB. 
These surfaces can have one or more regions of closed current surfaces detached from the wall, \textit{i.e.}~from the 
helical (primary) electrodes, by a separatrix with an X-point or by a tangency at the wall. 
The current steamlines lying on
closed $g$ surfaces around a maximum or minimum of $g$ correspond to pure secondary-to-secondary transformer current. 
The relative importance of the terms contributing to the evolution of $g$ is estimated.
While the contours of $g$ determine surfaces on which current density streamlines lie, they do not determine how far in $z$ the streamlines extend on these surfaces. This issue is addressed by metric (II), described in Sec.~\ref{sec:Delta-z}, which 
focuses on the axial displacement of the current streamlines that originate from the transformer primary at $r=r_w,z=z_0$ and terminate at $r=r_w,z=z_{final}$, namely
$\Delta z=z_{final}-z_0$. 
This tool determines whether or not there is a direct connection via current density lines
between the primary and secondary electrodes:
Current lines which exceed the periodicity length ($|\Delta z|/L>1$) represent, in a qualitative manner because of the 
periodic geometry we use, current that flows
directly (shunts) from the primary electrodes to the secondary electrodes, while current lines with $|\Delta z|/L<1$ suggest no shunting. 
Closed current streamlines inside a separatrix or tangency do not intersect $r=r_w$ and are considered to 
have infinite $\Delta z$. In the case of a separatrix, $\Delta z$ in fact approaches infinity as the separatrix
is approached from the outside.

A detailed sensitivity analysis to velocity boundary conditions and various resistivity profiles is conducted for the steady state. 
Four velocity boundary conditions (VBC) are employed. 
The first choice of VBC, designated $E \times B$, has the radial velocity $v_r$ set to the $(\mathbf{E}\times \mathbf{B}/B^2)_r$ drift for the $(1,1)$ and $(-1,-1)$ components, with the remaining Fourier components, and all of the Fourier components of $v_{\theta}$ and $v_z$, set to zero. 
The second VBC consists of no-slip (NS), or homogeneous Dirichlet, condition imposed on all components: $v_r$, $v_{\theta}$ and $v_z$. 
The third choice consists of homogeneous Neumann (HN) condition, applied again to all three components. 
This condition on the two tangential components corresponds to a zero-stress condition, as explained in the Appendix. 
The fourth VBC consists of zero-stress (ZS) condition for the 
two tangential components together with $v_r$ specified as the $E \times B$ condition 
as in the first set of conditions. 
The first condition was used for the nominal state of AFNB. 
The second and third conditions were briefly discussed in AFNB. 
The fourth condition, (zero-stress), provides a choice that is close to the natural radial velocity at the wall (see the Appendix).
It is found that the major differences in the results among the four sets of VBC relate to the helical field $g$ and the current line length $\Delta z$, \textit{i.e.}~metrics (I) and (II) above. Other quantities are affected by the VBC, but not in a qualitatively important manner.

The findings indicate that for sufficiently large $\phi_0$, $E\times B$ boundary conditions produce closed $g$ 
surfaces with a tangency while homogeneous Neumann and 
zero-stress conditions -- which behave very similarly -- produce closed $g$ surfaces with separatrices. 
For either case, the volume of the closed $g$ surfaces, and hence the amount of secondary current, grows 
with the magnitude of the applied potential $\phi_0$. 
No closed $g$ surfaces are observed with no-slip conditions under any circumstances. 
The current streamline diagnostic indicates that the cases that exhibit closed $g$ surfaces with separatrices, 
\textit{i.e.}, homogeneous Neumann and zero-stress give rise to the largest displacement and $|\Delta z|/L>1$, suggesting  
shunting, while no-slip conditions always yield $|\Delta z|/L<1$. 
An explanation for these results is given in Sec.~\ref{sec:streams}.

In AFNB, two radial profiles of resistivity were studied. The first is a hollow profile with a maximum
at the edge and rising over a small 
radial extent, with $\eta(r_w)/\eta(0)=100$. The second is uniform or flat resistivity profile. Here, 
we introduce a third profile, still maximum at the edge, with $\eta(r_w)/\eta(0)=100$,
but which varies over a longer length scale, and such produces a \emph{diffuse} resistivity profile.
The changes that the main features undergo are compared and contrasted for the three different resistivity profiles. 
The flat profile leads to increased distortion of helical flux relative to that of the hollow resistivity profile, 
a larger radial displacement of the $O-$point (bigger $r_O$), 
a flatter $q_0(r)$, and nearly a cancellation of the total axial current $I_z\simeq0$. 
The slotting of the secondary electrode proposed in AFNB addresses this tendency for cancellation, or near cancellation, for all resistivity profiles. 
The diffuse profile leads to only modest shift/distortion of flux surfaces relative to the hollow profile, with a smaller value of $r_0$ and a larger gradient in $q_0(r)$ except near $r=0$. 
The velocity boundary conditions affect the dynamics in the same manner for each case regardless of the resistivity profile. 
The current streamline displacement exhibits some sensitivity to the resistivity profile for the diffuse profile simply because the increased effective resistivity lowers the plasma current density $j_z$, thereby shortening the length of the primary current streamlines.

A third and final sensitivity study focuses on the departure from the sinusoidal (``smeared'') 
electrostatic drive employed in AFNB 
(and in earlier sections in this paper) to one that models concentrated primary electrodes, 
electrodes which are more localized both 
in $\th$ and in $z$.
We find that nearly all of the previously observed qualitative behavior that pertains to the smeared electrode configuration, 
including the response to the velocity boundary conditions and resistivity profiles, remains unchanged for the simulations 
run with a concentrated-electrode configuration. 

This paper is organized as follows: Sec.~\ref{sec:Model} introduces the zero $\beta$ resistive MHD model including 
normalizations, the boundary conditions, and the profile of the resistivity. Section \ref{sec:Transient} begins with a 
treatment of the transient that precedes a strongly driven steady helical state, focusing on the magnetic helicity and 
the channels that dissipate the input power. 
This is followed in Sec.~\ref{sec:Pitch} by studies of the properties of the helical steady 
state related to the rotational transform or twist and writhe of magnetic field lines that lie on surfaces of constant helical flux $\chi$. 
Section \ref{sec:CD-streamlines} introduces the two metrics: (i) closed contours of the helical field $g$ indicative of a pure secondary plasma current and contributions to the component of Ohm's law responsible for the evolution of $g$, and (ii) the axial displacement of the primary current streamlines. These metrics are employed to assess the 
possibility of shunting. 
Sensitivity of the characteristics uncovered in AFNB to the velocity boundary conditions and various resistivity 
profiles are presented in Secs.~\ref{sec:VelBC} and \ref{sec:eta_profiles}. 
Section~\ref{concentrate} covers the effects of a concentrated-electrode drive on the results. 
Finally, the summary, conclusions, and possible future work are presented in Sec.~\ref{sec:Summary}.
Details related to the boundary conditions are presented in the Appendix.


\section{Computational model\label{sec:Model}}

The resistive MHD model was described in detail in AFNB and is summarized in this section. 
We assume zero plasma pressure and a constant and uniform plasma density.
This leaves us with a system comprising the equation of motion, 
the resistive
Ohm's law, and Faraday's law: 
\begin{equation}
\rho_{0}\left(\frac{\partial\mathbf{v}}{\partial t}+\mathbf{v}\cdot\nabla\mathbf{v}\right)=
\mathbf{j}\times\mathbf{B}+\mu\nabla^{2}\mathbf{v},\label{eq:Momentum}
\end{equation}
\begin{equation}
\mathbf{E}+\mathbf{v}\times\mathbf{B}=\eta\mathbf{j},\label{eq:Ohm}
\end{equation}
\begin{equation}
\frac{\partial\mathbf{B}}{\partial t}=-\nabla\times\mathbf{E},\label{eq:Faraday}
\end{equation}
where $\Bvec$ and $\Evec$ are the magnetic and electric fields, 
$\mathbf{j}=\nabla\times\mathbf{B}$ is the current density, $\rho_0$ is the (constant) plasma density, and $\mf{v}$ 
is the plasma flow velocity. 
The quantities $\eta$ and $\mu$ are the plasma resistivity and viscosity, respectively, and we define the kinematic 
viscosity $\nu=\mu/\rho_0$.
The geometry is a periodic cylinder, occupying $0\le r\le r_w$ and $0\le z \le L=2\pi R$, where $R$ is the major radius. 

The above equations have been non-dimensionalized by scaling lengths to the wall radius, $r_w$, the magnetic 
field $\mathbf{B}$ to $B_z(t=0)=B_0$, 
and time to the nominal Alfv\'{e}n time $\tau_{A}=r_w/v_{A}$, based on $B_{0}$ and $\rho_{0}$.
The velocity in these units is relative to $v_{A}$, and $\rho_0$ equals unity. 
The plasma resistivity can have a spatial variation $\eta=\eta(r)$, while the kinematic viscosity, $\nu$, is kept spatially uniform. 

The viscous term used in Eq.~(\ref{eq:Momentum}) follows from the symmetric
stress tensor $\Pi_s=\mu\left(\nabla 
\mathbf{v}+(\nabla \mathbf{v})^T \right)$, assuming $\nabla \cdot \mathbf{v}=0$. With this assumption, the stress tensor 
becomes $\Pi=\mu\nabla 
\mathbf{v}$, as discussed in 
Landau and Lifschitz\cite{LandauLifshitz}. However, note that the condition $\nabla \cdot \mathbf{v}=0$ is
not enforced in DEBS and in fact is often not very small. (We will return to this issue in 
Sec.~\ref{sec:VelBC}, where we find that compression can affect the surfaces of the helical field $g$.) 
The Lundquist number is defined as $S=\tau_{R}/\tau_{A}$ where $\tau_{R}=r_{w}^{2}/\eta(r=0)$
is the resistive diffusion time. 
The Reynolds number is $Re=\tau_{v}/\tau_{A}=1/\nu$, where $\tau_{v}=r_{w}^{2}/\nu$ is the viscous diffusion time.

Equations (\ref{eq:Momentum})-(\ref{eq:Faraday}) are advanced with the 
DEBS code\cite{DEBS} and have been benchmarked for several cases with the 
NIMROD code\cite{NIMRODcode} in a periodic geometry. 
DEBS advances the vector potential $\mathbf{A}(\mathbf{x},t)$ rather than the magnetic field $\mathbf{B}(\mathbf{x},t)$. 
It uses the Weyl or temporal gauge\cite{Jackson2002,Akcay2017A} $\phi=0$, with $\mathbf{E}=-\partial\mathbf{A}/\partial t$, 
and in 
steady state the electric field is related to the part of $\mathbf{A}$ that is proportional to time.
For the spatial discretization, DEBS uses a finite difference approximation for the radial variation and a Fourier 
representation 
for the variation of the fields in the poloidal and 
axial directions: \textit{e.g.}~$\mathbf{E}(\mathbf{r})=\mathbf{E}(r)e^{im'\theta+ik'z}=\mathbf{E}(r)e^{im'\theta-in'\zeta}$, 
where $(m',n')$ represent the entire Fourier spectrum used in DEBS, including the driven mode $(m,n)=(1,1)$, the toroidal angle is $\zeta=z/R$, and $k'=-n'/R$. 

The zero-stress boundary conditions on the velocity are related to the exact form of the stress tensor $\Pi=\mu\nabla 
\mathbf{v}$, which leads to the viscous operator in Eq.~(\ref{eq:Momentum}). This is discussed in detail in the Appendix.

The boundary conditions on the fields are prescribed in terms of a voltage $\phi_{0}$ at $r=r_w=1$ applied to
the Fourier harmonic with $(m,n)=(1,1)$, where $(m,n)$ (and its complex conjugate which is always implied wherever the $(1,1)$ pair appears in this paper) designate the particular pair(s) of harmonics that is(are) driven:
\begin{align}
\label{eq:ESBC-1}
E_{\theta}^{(m,n)}(1) &=-(im/r_{w})\phi_{wall} = -i\phi_{wall}/r_w\\
 E_{z}^{(m,n)}(1) &=-ik\phi_{wall}= i\phi_{wall}/R\label{eq:ESBC-2}
\end{align}
where $k=-n/R=-1/R$ and $\phi_{wall}=\phi_{0}e^{i\theta-iz/R}$ for $(m,n)=(1,1)$. 
Also, we have $E_{\theta}^{-(m,n)}(1)=E_{\theta}^{(m,n)*}(1)$ and similarly for $E_{z}$. 
These relations result in 
\begin{equation}
mE_{z}^{(m,n)}-krE_{\theta}^{(m,n)}=E_{z}^{(1,1)}+\frac{r}{R}E_{\theta}^{(1,1)}=0.
\label{eq:Electrostatic-BC}
\end{equation}
For all other Fourier components, the tangential
components of the electric field are zero at the wall; these conditions are consistent with 
$(\partial/\partial t) B_{r}^{(m,n)}=0$ for all $(m,n)$. 
We also specify $A_{\theta}$ and $A_z$ at $t=0$ such that $B_{r}^{(m,n)}=0$ for all Fourier harmonics;
in spite of this, the conditions in Eqs.~(\ref{eq:ESBC-1}) and (\ref{eq:ESBC-2}) on the tangential field, 
$\mathbf{E}_t=-\nabla_t \phi_{wall}$,
is weaker than the commonly used perfectly conducting conditions $E_{\theta}^{(m,n)}=E_{z}^{(m,n)}=0$ for all $(m,n)$. 
An alternate formulation of the EM boundary conditions in terms of a normal current density $j_{r}^{(1,1)}$ at $r=1$ was presented in AFNB and is used in Sec.~\ref{concentrate} to impose the helical drive \textit{via} much more localized (concentrated) electrodes. 
The single harmonic description of AFNB differs from the multi-harmonic implementation in Sec.~\ref{concentrate} in that the former prescribes only $j_r^{(1,1)}$ with zero tangential electric field for $(m',n')\ne(1,1)$, while the latter prescribes all Fourier components $j_r^{(m',n')}$. 

We apply four sets of velocity boundary conditions.
The first VBC consists of $E\times B$ boundary conditions on the radial component of the
velocity at $r=1$ for the $(1,1)$ Fourier amplitudes and the no-slip condition on the remaining Fourier components of $v_r$ as well as on the tangential 
components ($v_{\th}(1)=v_z(1)=0$).  
Specifically, we take $v_{r}^{(1,1)}(1)=\hat{\mathbf{r}}\cdot\mathbf{E}^{(1,1)}
\times\mathbf{B}^{(0,0)}/(B^{(0,0)})^{2}$, in other words,
linearizing with respect to the $(1,1)$ component of the applied electric field, and 
similarly for $(-1,-1)$.
The Appendix shows estimates of the relative contribution of $\eta \jvec_{\perp}\times \mathbf{B}/B^2 \text{ to } 
\mathbf{E}\times \mathbf{B}/B^2$ and its dependence on plasma parameters.
The second VBC consists a of no-slip (homogeneous Dirichlet) conditions $\mathbf{v}=0$ at $r=r_w$. 
The third VBC consists of homogeneous Neumann conditions $\partial \mathbf{v}(r=1) / \partial r=0$, which are zero-stress conditions when applied to the tangential components. This issue will be discussed in more detain in Sec.~\ref{sec:VelBC}.
The fourth set imposes a zero-stress condition on the tangential velocity components while setting 
$v_r(1)$ to equal the radial component of the $E\times B$ drift, as in the first set of VBC. 
Henceforth, the first VBC will be referred to simply as $E\times B$, the second as no-slip (NS), the third as homogeneous Neumann (HN), and the fourth as zero-stress (ZS).

The resistivity is specified as a function of radius to be of the form 
\begin{equation}
\eta(r)=\eta(0)[1+(\sqrt{\eta(1)/\eta(0)}-1)r^{p}]^{2}.\label{Eq:resistivityprofile}
\end{equation}
A ``hollow'' profile ($p=16$) and a ``diffuse'' profile $p=4$, both with $\eta(1) / \eta(0)=100$ as well as a flat profile with $\eta(1)=\eta(0)$ are studied. 
The first and third profiles were used in AFNB. 
For $\eta(1)>>\eta(0)$, the hollow and diffuse profiles provide a
large resistivity near the wall, with $\eta(r)\approx\eta(0)$ in an inner region, which spans most of the interior for the hollow case and is smaller
for the diffuse case. The high edge resistivity can be thought of as a model for sheaths around the helical electrodes.

\section{Further characterization of the nominal state\label{MoreStuff}}
In this section we extend the helical diagnostics of AFNB to the transient stage and 
also describe the evolution of the magnetic 
helicity and power balance. 
The field line pitch on the actual helical flux surfaces appropriate for helical symmetry is also 
presented in this section.
\subsection{The transient stage; magnetic helicity and energy\label{sec:Transient}}
\begin{SCfigure*}
\centering
\includegraphics[width=0.75\textwidth]{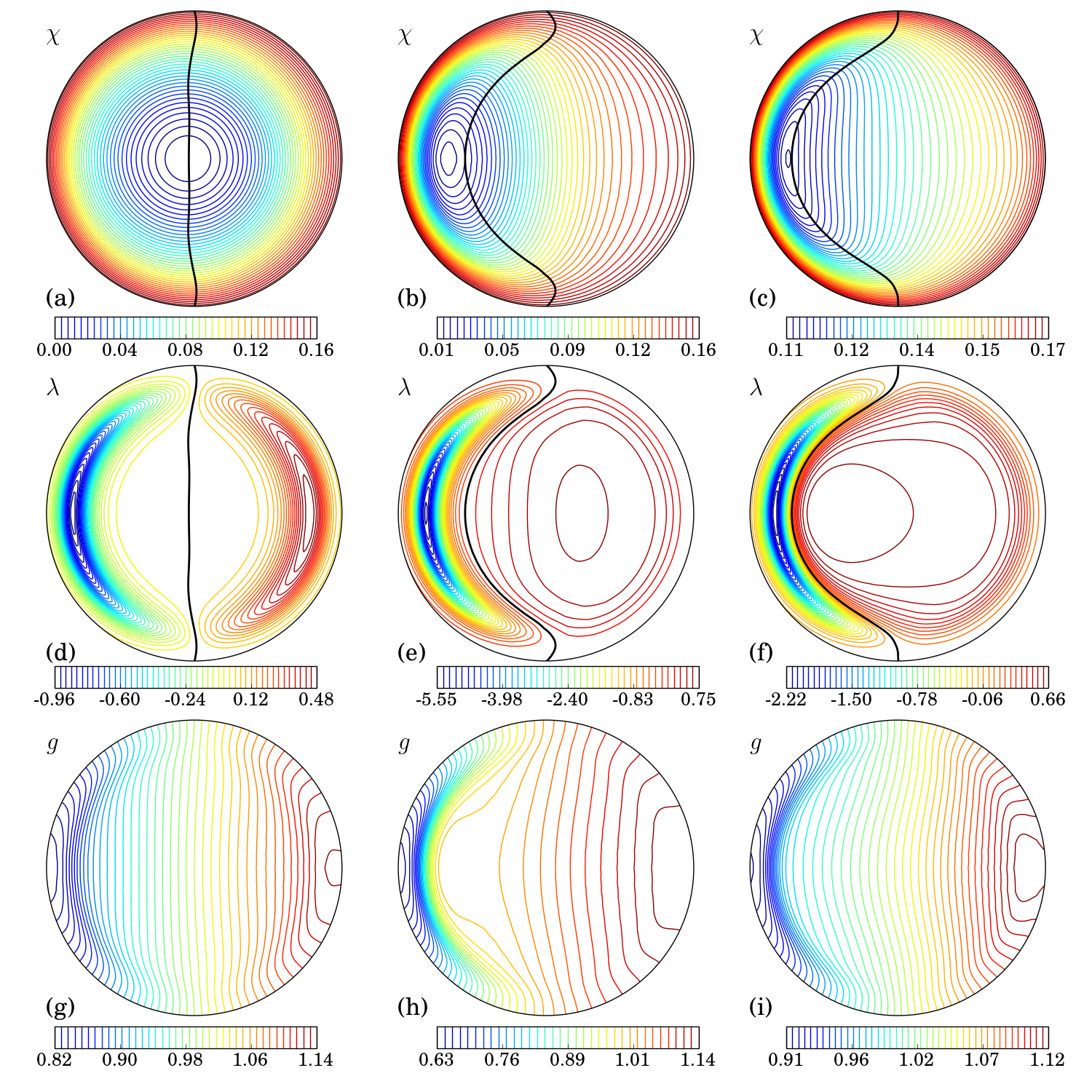}
\caption{\label{fig:transient} Evolution of the helical flux $\chi$ (top row), parallel current density $\lambda$ (middle row), 
and helical field $g$ 
(bottom row) during the transient stage that leads 
to the time-asymptotic state with a flat $q_0(r)\gtrsim 1.0$ in the plasma interior. Parameters are identical to those of the 
nominal state in AFNB. 
The plotted quantities are, from top to bottom, 
for time $t/\tau_{A}=1.5, 5.3, 11$, respectively. The helical perturbation is ramped up over a period of $\tau_{ramp}=2.5\Talf$. 
The contour $\lambda=0$ 
(black) is also superposed in (a)-(f).}
\end{SCfigure*}

The transient state described here is for the nominal case of AFNB, which is driven with a $(m,n)=(1,1)$ 
helical electrostatic potential of 
amplitude $\phi_0=0.2$, ramped up over a period of $\tau_{ramp}=2.5\Talf$. The VBC are the $E\times B$ conditions, and 
the resistivity profile is the hollow profile. The Lundquist number is $S=100$ and the Reynolds number is $Re=10$.
For a single helicity state, the helical flux $\chi(r,u)\equiv r\boldsymbol{\sigma}\cdot\mathbf{A}$ and helical 
field $g(r,u)\equiv r\boldsymbol{\sigma}\cdot\mathbf{B}$, 
where $\boldsymbol{\sigma}=\hat{\mathbf{r}}\times \mathbf{k}$ and $\mathbf{k}=\nabla u=m\hat{\mathbf{\theta}}/r+k 
\hat{\mathbf{z}}$,  become
\begin{align}
\chi(r,u) &= mA_{z}(r,u)-krA_{\theta}(r,u)\nonumber\\
          &= A_{z}(r,u)+rA_{\theta}(r,u)/R,\label{eq:HelicalFlux}\\
g(r,u) &=mB_{z}(r,u)-krB_{\theta}(r,u)\nonumber\\
       &= B_z(r,u) + rB_{\th}(r,u)/R,\label{Eq:HelicalField}
\end{align}
where $m=n=1$ is substituted into the latter forms. In terms of $\chi$ and $g$, the helical representations 
for the magnetic field 
and the current density are
\begin{align} 
\mathbf{B}&=f(r) \nabla \chi(r,u) \times \boldsymbol{\sigma}+f(r)g(r,u)\boldsymbol{\sigma},\label{eq:B-representation}\\
\mathbf{j}&=f(r) \nabla g(r,u) \times \boldsymbol{\sigma}+f(r)h(r,u)\boldsymbol{\sigma},\label{Eq:j-representation}
\end{align}
where $f(r)=1/r|\mathbf{k}|^2$ and the occurrence of $g$ in the second term in Eq.~(\ref{eq:B-representation})
and in the first term 
in Eq.~(\ref{Eq:j-representation}) follows from $\mf{j}=\nabla \times \mf{B}$.

The helical flux $\chi$ is gauge invariant and proportional to the magnetic flux through
a ribbon with $u$ constant. 
The helical flux satisfies $\mathbf{B}\cdot\nabla\chi=0$, meaning that the $\chi$ surfaces are magnetic surfaces. 
Analogous to this, $g(r,u)$ satisfies $\mathbf{j}\cdot\nabla g=0$, implying that current density
streamlines lie on $g$ surfaces, and $g$ is proportional to the current through a ribbon with $u$ constant. Therefore,
if $g_{max}$ and $g_{min}$ are, respectively, the maxima and minima on $r=r_w$, then $\Delta g=g_{max}-g_{min}$
is proportional to the net current entering through the electrodes at $r=r_w$.
Note that for $q_0 \approx 1$, we have $rB_{z}/RB_{\theta} \approx 1$, so that  $g=B_z\left(1+r^2/R^2 q_0\right) =
B_z\left(1+O(\epsilon^2)\right)$, and thus $g\simeq B_z\simeq 1$. 

Two more observations follow from the representation in Eq.~(\ref{Eq:j-representation}). First, the current density satisfies
$\boldsymbol{\sigma}\cdot\mathbf{j}\times\mathbf{B}\propto\boldsymbol{\sigma}\cdot\nabla\chi\times\nabla g$,
so that the $g$ surfaces and the $\chi$ surfaces must align where
the plasma is force-free. The second observation relates to the fact that the parallel current density $\lambda$ is zero along
the O-line in steady state with zero back EMF. This point is mentioned in the introduction and is discussed at length in
AFNB, where it is discussed that this property cannot hold in axial symmetry, and that the rotational transform
along the O-line is due to helical geometry. 

For all the cases studied in this paper, the time-asymptotic state is steady-state
for $\phi_0 < \phi_{crit}$, above which a time-dependent state occurs. As in AFNB, we see no evidence of hysteresis, 
i.e.~multiple solutions.
Figure \ref{fig:transient} shows the evolution of the helical flux $\chi$ (top row, (a)-(c)), the parallel current density 
$\lambda=\mathbf{j\cdot B}/B^2$ (middle row, (d)-(f)), 
and the helical field $g$ (bottom row, (g)-(i)) at 
three different times, $t/\Talf = 1.5$, 5.3, and 11 during the transient stage. 
At $t/\tau_{A}=1.5$, well within $2.5\Talf$-long ramp-up period, the $O-$point in $\chi$ in 
Fig.~\ref{fig:transient}a 
appears to be very close to the $\lambda=0$ curve, 
which nearly intersects the origin this early in time.
As discussed in AFNB, the flux surface average $\langle \eta \lambda B^{2} \rangle=0$ on each flux surface in steady 
state if the back EMF $E_0$ is zero. The position of the $\lambda=0$ curve and the near circularity of 
the $\chi=$const.~surfaces are consistent with the very small value of the flux surface 
average at this early stage. 
During this stage, the poloidal field $B_{\theta}$ and therefore $A_z$ are very small, leading to $\chi\simeq (r/R)A_{\theta}$. 
Since $r A_{\theta} = \int^{r}_{0} B_z(r')r' dr'= \Phi(r)/2\pi\approx B_0 r^2/2$, where $\Phi$ is the axial flux, the helical 
flux $\chi\approx \Phi/R$ 
is seen to have this behavior in  Fig.~\ref{fig:transient}a. In  Fig.~\ref{fig:transient}b, 
the $\chi$ surfaces are much more helically kinked and the $O-$point is noticeably to the left of the $\lambda=0$ curve,
similar to results shown in AFNB with back EMF (Fig.~8). Consistent with this point, the 
flux surface 
average $\langle \eta \lambda B^2 \rangle$ on 
the inner surfaces near the $O-$point is negative. 
By the time of Fig.~\ref{fig:transient}c, right before steady state, 
the $O-$point once again approaches 
the $\lambda=0$ curve, with a very small $\langle \eta \lambda B^2 \rangle$, as discussed in AFNB.

\begin{figure}
\centering\includegraphics[width=0.5\textwidth]{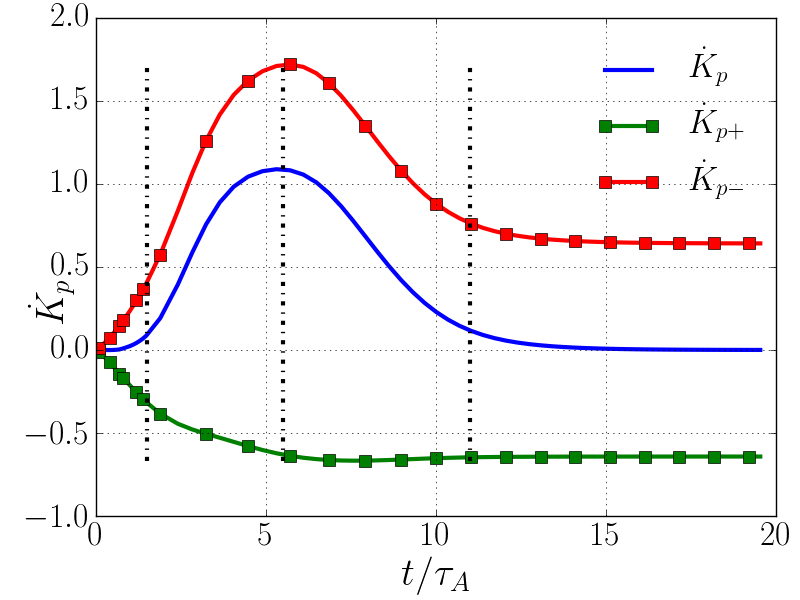}
\caption{\label{fig:Kdot_vs_time} The total rate of change of magnetic helicity $\dot{K}_p$, as well as $\dot{K}_{p-}$ and $\dot{K}_{p+}$, the rate of change of helicity in the negative $\lambda$ and positive $\lambda$ regions, respectively, as functions of normalized time $t/\Talf$. 
The vertical dashed-dotted lines correspond to the three times chosen for Figure \ref{fig:transient}.}
\end{figure}

\begin{figure}
\centering
\includegraphics[width=0.5\textwidth]{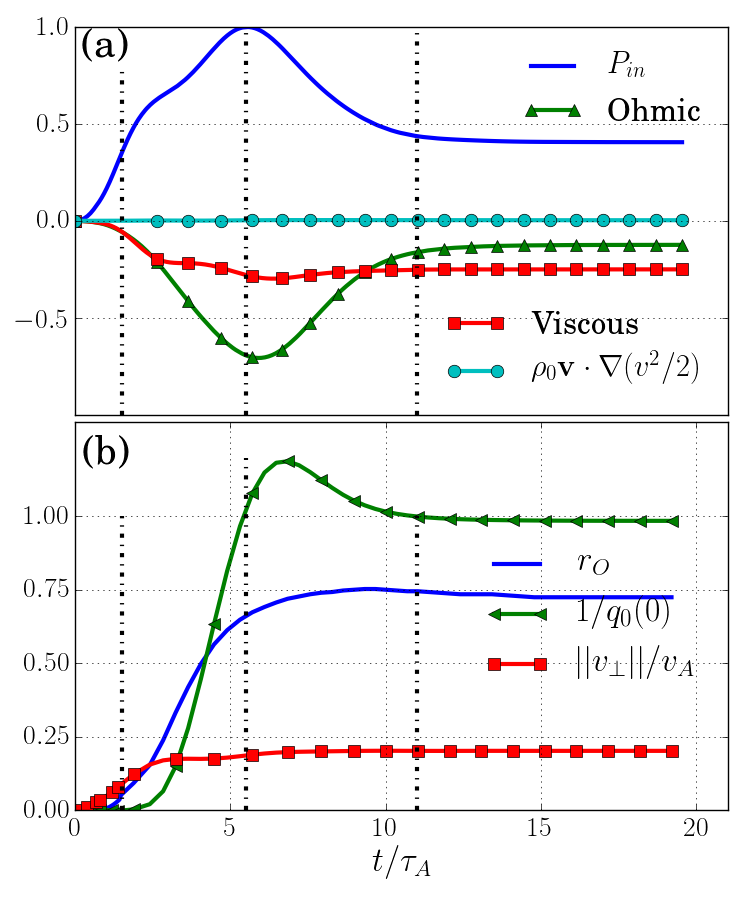}
\caption{\label{fig:power_vs_time} Terms (a) in the power balance as functions of normalized time $t/\Talf$: the input 
Poynting flux, 
the Ohmic dissipation and the viscous dissipation. Also 
shown is the small term due to not evolving the density via the continuity equation. In (b) are the norm of the perpendicular 
velocity 
(in Alfv\'{e}n units), the position of the $O$-point, and the reciprocal of the mean safety factor on axis $1/q_0(0)$. 
The vertical dashed-dotted lines correspond again 
to the three times chosen for Fig.~\ref{fig:transient}. }
\end{figure}

The parallel current in Fig.~\ref{fig:transient}d appears to have a nearly a pure $m=1$ like structure during the early 
stage, as 
expected based on linearity with respect to $\phi_0$ at this stage. 
At the two later times (Figs.~\ref{fig:transient}e, f), the strong helical perturbation and a $(0,0)$ component are 
evident.
At $t/\Talf=5.3$ the $\lambda<0$ region attains values that are several times larger in magnitude than the value at 
steady state. 
This is around the time of the peak power injection and helicity generation (Figs.~\ref{fig:Kdot_vs_time} and 
\ref{fig:power_vs_time}).

The helical field $g$ is close to $B_0=1$, with contours (current density surfaces) that show the injection and exiting
of electrode current 
$\propto \Delta g=g_{max}-g_{min}$ at the top and bottom, respectively during the early (Fig.~\ref{fig:transient}g, $\Delta g=0.32$) 
and late 
transient periods (Fig.~\ref{fig:transient}i, $\Delta g=0.21$). 
When the transient activity is at strongest, in Fig.~\ref{fig:transient}h, the net current is larger, $\Delta g=0.51$. 
Figures \ref{fig:transient}g-i appear to have very small areas of closed $g$ contours which appear to separated 
from the open $g$ surfaces by a tangency
at the wall. 
The amount of current contained in this closed current line region (detached from the electrodes) is quite small. 
The surfaces of $g$ also exhibit in the left half of Fig.~\ref{fig:transient}h a second region of weak closure 
with a separatrix, at the time of peak helicity/power injection. (See Figs.~\ref{fig:Kdot_vs_time} and 
\ref{fig:power_vs_time}.)
The amount of net parallel current in this latter closure region appears to be small since this 
separatrix overlaps regions of both $\lambda<0$ and $\lambda>0$ (Fig.~\ref{fig:transient}e).

Fig.~\ref{fig:Kdot_vs_time} shows the total rate of change of helicity $\dot{K}_p$ (solid blue) as well as $\dot{K}_{p+}$, 
the contribution to  $\dot{K}_p$ from 
the $\lambda>0$ region (green triangles), 
and $\dot{K}_{p-}$, the contribution to $\dot{K}_p$ from the $\lambda<0$ region (red squares). As discussed in AFNB, $\dot{K}_{inj}=0$ , since $B_r(r=r_w)=0$. 
This figure shows that very early in time, when the imposed perturbation grows linearly in the plasma, the 
quantities $\dot{K}_{p+}$ and $\dot{K}_{p-}$ are equal and opposite. The period of increase in magnetic helicity is 
$2 \lesssim t/\tau_A \lesssim 10$, with maximum of $\dot{K}_p$ and $\dot{K}_{p+}$ occurring near $t/\tau_{A}=5$. 
Consistently, the area with $\lambda<0$ has grown largest at this point, $\langle \eta \lambda B^2 \rangle<0$ 
 on the inner flux surfaces, and the $O-$point is farthest from the $\lambda=0$ surface. 
(See Fig.~\ref{fig:transient}b.) Near the end of the 
transient period (at e.g.~$t/\tau_{A}=11$), $\dot{K}_{p+}$ and $\dot{K}_{p-}$ have both become
constant and are equal and opposite, yielding $\dot{K}_{p}=0$, and consequently, the $O-$point migrates back to the $\lambda=0$ curve as shown in Fig.~\ref{fig:transient}c. It is 
clear that $\dot{K}_{p-}$ produces helicity while $\dot{K}_{p+}$ dissipates it, consistent with $\dot{K}_p=-2\int \eta \jvec 
\cdot \mathbf{B}$. 
That is, the region with $\lambda<0$ produces helicity 
and the region with $\lambda>0$ dissipates it. The positive and negative terms $\dot{K}_{p+}$ and $\dot{K}_{p-}$
balance in steady-state, but an excess is produced during the transient, so that the 
steady-state has positive helicity, independent of time. 
This particular way of driving the plasma suggests creating a region with $\lambda<0$ instead of employing 
electrostatic helicity injection  
\textit{via} a normal magnetic field and potential $\phi$ at the wall.


The power balance in zero pressure visco-resistive MHD with density $\rho=\rho_0$ constant, in a volume $V$ bounded 
by a surface $S$, 
for energy $E=\int_V dV (\rho_0 \mathbf{v}^2/2+\mathbf{B}^2/2)$, is as follows: 
\begin{align}
\frac{d E}{d t}&=-\int_{S}\hat{\mathbf{n}}\cdot\mathbf{E}\times\mathbf{B}dS- \int_V \eta \jvec^2dV\nonumber\\
             &+ \int_V \mu\vvec\cdot\nabla^2 \vvec dV-\int_V\rho_{0}\mathbf{v}\cdot\nabla
             \left(\frac{\mathbf{v}^{2}}{2}\right)dV.
\label{eq:dEdt}
\end{align}
The first term represents the Poynting flux through the surface $S$. For back EMF $E_0=0$ this equals
\begin{align}
P_{in}=-\int_{r=1} rd\theta dz(E_{\theta}B_z-E_z B_{\theta})\\
=\int g (\partial \phi /\partial u) d\theta dz
=-\int \phi (\partial g /\partial u) d\theta dz,
\end{align}
and is the input Poynting flux due to the helical electrodes. Also, the relation $j_r=(1/r)\partial g/\partial u$ from 
Eq.~(\ref{Eq:j-representation}) 
implies that $P_{in}$ equals 
$-\int \phi j_r rd\theta dz$. The results show that, 
indeed, $\phi$ and $g$ are out of phase 
by $90^o$ ($\phi$ and $j_r$ out of phase by $180^o$), giving maximum input power. (As discussed in AFNB, for $E_0\neq 0$ 
there is an Poynting flux term 
proportional to the back EMF $E_0 L$.) The second and third terms on the right in Eq.~(\ref{eq:dEdt}) 
are the Ohmic and viscous losses, respectively, and the 
last term is the due to the violation of energy conservation due to the assumption of constant density $\rho=\rho_0$, as discussed in AFNB.
These four quantities are traced as a function of the normalized time in Fig \ref{fig:power_vs_time}a. 
The input power $P_{in}$ (solid blue) rises until it reaches its peak at $t/\Talf=5$ in the first 
half of the 
transient stage, approximately where $\dot{K}_{p}$ is maximum in Fig.~\ref{fig:Kdot_vs_time}. All the terms appearing in 
the figure are scaled by 
the peak input power. 
During this period, $P_{in}$ is predominantly lost to Ohmic dissipation (green triangles), which dissipates approximately 
70\% of $P_{in}$ 
at its peak at $t/\Talf=5$. 
As the configuration nears its steady-state, however, the Ohmic power drops rapidly as the viscous dissipation overtakes 
it and becomes 
the main dissipation channel.  
Note the evolution of the viscous dissipation is closely correlated with that of the perpendicular plasma flow as shown 
by the history 
of the norm of the perpendicular velocity $||\mf{v}_{\perp}||/v_A$ (red squares) 
in Fig.~\ref{fig:power_vs_time}b. The norm here is defined by $||v_{\perp}||\equiv\int|\mathbf{v}_{\perp}|dV/\pi r_{w}^{2}L$.
The term associated with the lack of energy conservation due to a constant density assumption is negligible for all time. 
However, as was discussed in the Appendix of AFNB, this term is proportional to $(|\mathbf{v}_{\perp}|/v_A)^3\sim \phi_0^3$
and can become significant for $||\mf{v}_{\perp}||/v_A\rightarrow 1$. 
Also shown in Fig.~\ref{fig:power_vs_time}b are the histories of the radial position of the $O-$point, $r_O$ (solid blue) 
and the reciprocal of the mean 
safety factor on axis, $1/q_0(0)$ (green). The former converges to its-steady state value 0.73 at $t/\Talf=10$. 
Note that there is a small overshoot in $1/q_0(r=0)$, which slightly lags the peaks in $\dot{K}_{p+}$ and $P_{in}$\cite{FootNote1}.
This behavior is less evident for larger values of $\tau_{ramp}$.
The rate of change of the total energy converges to zero in steady-state, for $t/\tau_A \gtrsim 10$.

\subsection{Helical field line pitch in steady state\label{sec:Pitch}}

Let us define $q_{h}(\chi)$ as $\Delta z/2\pi R$, where $\Delta z$
is the change in $z$ following the magnetic field lines over one
circuit around the $\chi=$ const.~surface in the helical variable
$u=m\theta+kz$. We write $dz/du=B_{z}/\mathbf{B}\cdot\nabla u$,
leading to 
\begin{equation}
q_{h}(\chi)=\frac{1}{2\pi R}\ointop\frac{B_{z}du}{\mathbf{B}\cdot\nabla u}.\label{eq:q-by-field-lines}
\end{equation}
This quantity is a function of $\chi$ because the field line integration
is on flux surfaces. As before, we specialize here to $(m,n)=(1,1)$. The coordinates
are $(r,u,z)$ with a nonorthogonal covariant basis $\nabla r$, $\nabla u$, $\nabla z$
and the integral over $u$ in Eq.~(\ref{eq:q-by-field-lines}) is
from $u_{min}(\chi)$ to $u_{max}(\chi)$. For steady state with $m=1$ the magnetic axis in the
presence of a helical perturbation is displaced from the axis
$r=0$ as in Figs.~\ref{fig:transient}b,c, so we must distinguish between $\chi$ surfaces which encircle
the axis $r=0$ and those that do not. For the former we have $u_{min}=-\pi$,
$u_{max}=\pi$\cite{FootNote2}. 
We can express $q_{h}(\chi)$ in terms of fluxes by first writing the physical toroidal
flux $\Phi_p=2\pi\Phi$ within an area encircled by $\chi=\mbox{ const.}$~as

\begin{equation}
\Phi_{p}(\chi)=\int_{\chi'\leq\chi}B_{z}dS=\int B_{z}\frac{d\chi' du}{|\nabla\chi'\times\nabla u\cdot\hat{\mathbf{z}}|}.
\label{eq:Toroidal-flux}
\end{equation}
The limits on $u$ are as in Eq.~(\ref{eq:q-by-field-lines}) discussed
above. From the helical representation in Eq.~(\ref{eq:B-representation}) and from
the observation $du/\mathbf{B}\cdot\nabla u<0$, we find 
$\nabla\chi\times\nabla u\cdot\hat{\mathbf{z}}=-r\boldsymbol{\sigma}\cdot\hat{\mathbf{z}}\mathbf{B}\cdot\nabla u$
or, for $m=1$,
\[
|\nabla\chi\times\nabla u\cdot\hat{\mathbf{z}}|=-\mathbf{B}\cdot\nabla u.
\]

From these considerations we conclude
\begin{equation}
q_{h}(\chi)=-\frac{d\Phi_{p}}{d\chi_{p}},\label{eq:FluxDerivatives}
\end{equation}
where the physical helical flux is $\chi_{p}=2\pi R\chi$. Numerically, we compute the toroidal
flux $\int B_z dS$ within each $\chi$ surface and differentiate with respect
to $\chi$. This method, which requires some smoothing for data on
a grid, e.g.~$\chi_{i,j}=\chi(r_{i},u_{j})$, obviates the need for
field line integrations and the complications regarding $\chi$ surfaces
which encircle $r=0$ and those that do not. 

\begin{figure}
\centering
\includegraphics[width=0.5\textwidth]{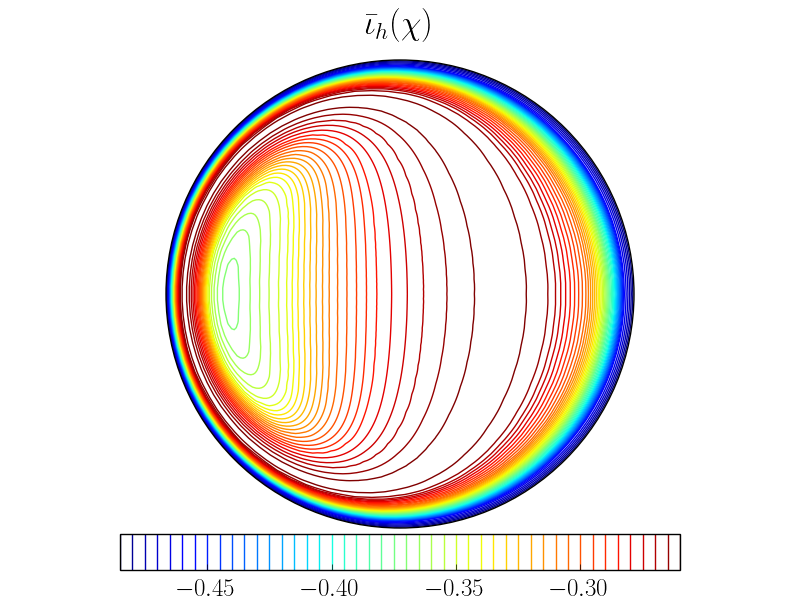}
\caption{\label{fig:iota_h} Contours of the helical transform $\bar{\iota}_{h}(\chi)=-d\chi_{p}/d\Phi_{p}$ for the nominal case of AFNB. }
\end{figure}

If we average the expression $\chi=mA_{z}-krA_{\theta}=\psi+\Phi/R$
(for $m=n=1$) over the $\chi$ surfaces and multiply by $2\pi R$,
we obtain
\[
\chi_{p}=\psi_{p}+\Phi_{p},
\]
where $\psi_{p}(\chi)=2\pi R\langle A_z \rangle$. This leads to 
\begin{equation}
\bar{\iota}_{h}(\chi)=\bar{\iota}(\chi)-1,\label{eq:iota-bar-h}
\end{equation}
where the helical transform is $\bar{\iota}_{h}=1/q_{h}=-d\chi_{p}/d\Phi_{p}$ and the rotational 
transform is $\bar{\iota}=1/q=-d\psi_{p}/d\Phi_{p}$.
(The signs are consistent with $u=m\theta+kz=\theta-\zeta$, so that
$\Delta u=\Delta\theta-\Delta\zeta$.) As expected, the result in
Eq.~(\ref{eq:iota-bar-h}) shows that if $\bar{\iota}(\chi)=1/q(\chi)$ equals unity, we
have $\bar{\iota}_{h}(\chi)=0$, i.e.~that the field lines on the
surface labeled by $\chi$ rotate at exactly the same rate as the
magnetic axis $\chi=\chi_{min}$, which is $n/m=1$.  That is, the quantity
$\bar{\iota}_{h}(\chi)$ in Eq.~(\ref{eq:iota-bar-h}) gives the
twist of the field lines\cite{Bellan-textbook} on the $\chi=$const.~surfaces associated with the helical symmetry in addition to the writhe \cite{Bellan-textbook} $n/m=1$.

Figure~\ref{fig:iota_h} shows the contours of $\bar{\iota}_{h}(\chi)$ for the nominal case of AFNB.
These results show $\bar{\iota}_h<0$ or $q>1$, qualitatively consistent
with $q_{0}(r)>1$. The amount of twist, $\bar{\iota}_h \lesssim -0.25$, has a maximum on flux surfaces 
passing through $(r,\theta)=
(0.4,0)$. The maximum value 
of $\bar{\iota}_h$ is not as close to zero 
as one might expect, given that the mean field value $q_0(r=0)$ is very slightly above unity. This 
appears to be related to the 
fact that each flux surface encircles
the $O-$point, approximately at $(r,\theta)=(0.75,\pi)$ for the nominal case of AFNB shown, crossing 
through a region with higher $|\nabla \chi|$ 
and higher 
local twist (where the local field line twist rises sharply). Cases that are more strongly driven 
(larger $\phi_0$ or higher $S$) 
show the maximum of $\bar{\iota}_h$
converging to zero from below ($q(\chi)\rightarrow 1+$) and becoming flatter.

These results show that there is some twist on the $\chi$ surfaces in addition to the writhe $n/m=1$. 
Qualitatively speaking, one expects that this distribution between twist and writhe 
should minimize the energy in the magnetic field. 
The conclusion $\lambda_{O}=0$ for the steady state solutions with zero back EMF in AFNB shows that this distribution 
of twist and writhe must be consistent with having exactly zero parallel current density on the magnetic axis (as 
well as the flux surface average $\langle\eta\lambda B^{2}\rangle=0$ on each constant $\chi$ surface.) As discussed in AFNB, 
the property of having an O-point where the current density is zero is due to helical symmetry, and cannot occur for axial
symmetry ($n=0$) or azimuthal symmetry ($m=0$). Also, along the O-line $\nabla \chi$ is zero, so $\mathbf{B}$ is parallel
to $\boldsymbol{\sigma}$, $\lambda=0$ implies (c.f.~Eq.~(\ref{Eq:j-representation})) 
$\mathbf{j}\cdot \boldsymbol{\sigma} \propto \lambda$, showing that $h=0$ along the O-line.

\section{Helical field $g$ surfaces and current density streamlines\label{sec:CD-streamlines}}

This section introduces two issues that relate to the physics of this device. The first,
in Sec.~\ref{sec:g-field-terms}, is the 
evolution of the surfaces of the helical field $g$. 
The significance of having closed $g$ surfaces is that this suggests a pure 
secondary-to-secondary current disconnected from the primary electrodes. 
The second issue, discussed in Sec.~\ref{sec:Delta-z}, deals with  
the integration of the current streamline length $\Delta z$, 
which provides a quantitative determination of how far the primary current trajectories extend axially and 
an indication of whether or not direct shorting (shunting) occurs between the primary and secondary electrodes.

\subsection{Contributions to the helical field $g$\label{sec:g-field-terms}}


The results shown in Fig.~\ref{fig:transient} and AFNB show that closed surfaces of $g$, with 
enclosed current detached from the primary 
electrodes, can be present with $E\times B$ boundary conditions. The components of current
density proportional to $\nabla g \times \boldsymbol{\sigma}$ in Eq.~(\ref{eq:HelicalFlux}) 
run mainly from top to bottom, from one primary electrode to the other in Fig.~\ref{fig:transient}.
However, a small area in this figure indicates that a very 
small amount of current encircles 
the $O$-point in the $g$ surfaces in a counter-clockwise manner, where $g$ has its maximum value. These
closed surfaces are separated from the open surfaces by a tangency at $r=r_w$. They 
represent, within the approximation of modeling a finite length system with 
a geometry periodic in $z$, current that flows from one secondary electrode at $z=0$ to the other at $z=L$,
disconnected from the primary electrodes at $r=r_w=1$. 
A small area containing closed surfaces with a local maximum also appears during the transient shown in 
Fig.~\ref{fig:transient}h. 
However, these surfaces are separated from the surfaces of open streamlines by a separatrix with an X-point. 

\begin{figure}
\centering
\includegraphics[width=0.5\textwidth]{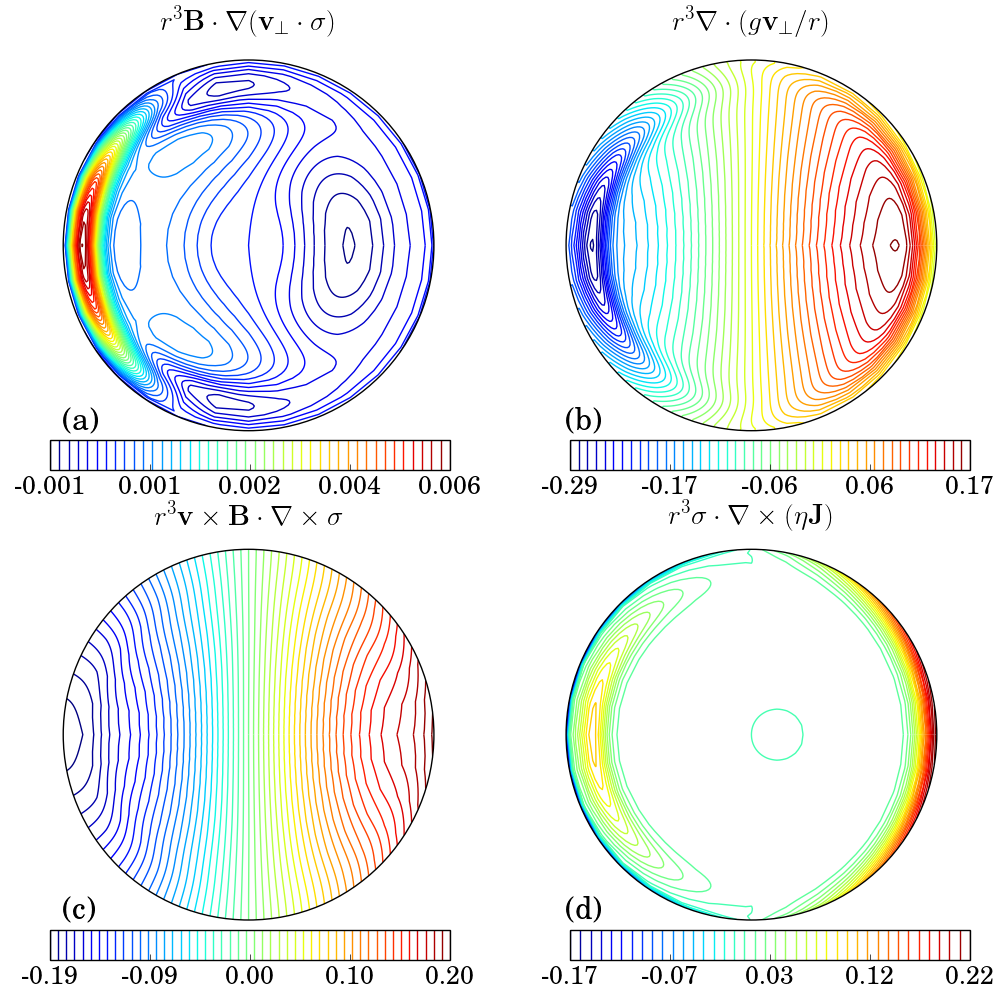}
\caption{\label{fig:g_evolution} The four quantities on the right of Eq.~(\ref{eq:g-advection-compression}) contributing to the evolution of the helical field $g$ ($g/r$) in steady state. 
These are, respectively, (a) the differential rotation, (b) the advection-compression, (c) $\nabla\times\boldsymbol{\sigma}$, and (d) the resistive terms.}
\end{figure}
\begin{SCfigure*}
\centering
\includegraphics[width=0.75\textwidth]{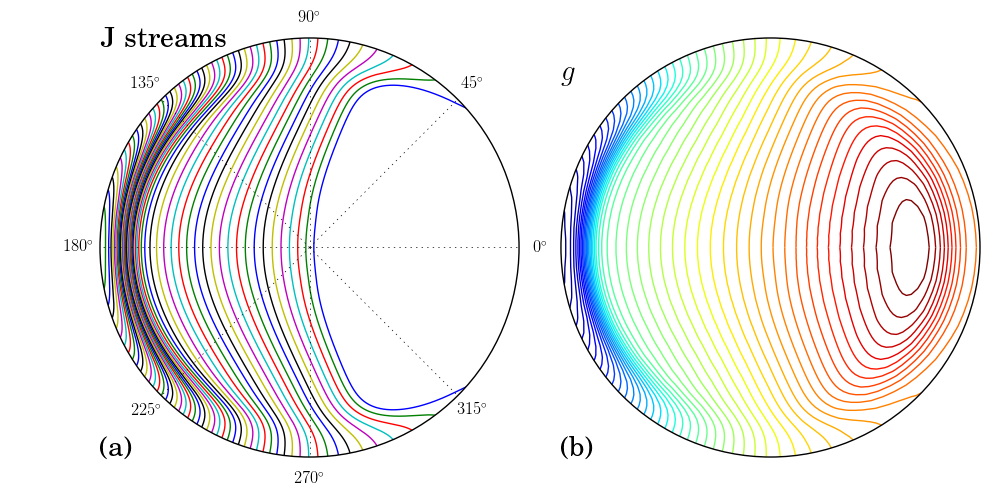}
\caption{\label{fig:J_streams} Orbits (a) of Eq.~(\ref{eq:Current-streamlines}) initialized at $r=r_w$ and 
values of $\theta$ with 
$j_r<0$,
from a strongly-driven simulation ($\phi_0=0.6$) that corresponds to the black trace of 
Fig.~\ref{fig:DeltaZ_vs_Uinit}. In 
(b) are shown the actual 
surfaces of the helical field $g$; the open $g$ surfaces match well with the streamlines of (a). 
Note that because the orbits in (a) are launched from the boundary, they cannot sample any point 
within the region of closed $g$ surfaces. }
\end{SCfigure*}

In order to explore the origin of this region of closed current lines, we use
the helical representation in Eq.~(\ref{eq:B-representation}),
Faraday's law and Ohm's law to obtain,
\[
\frac{\partial}{\partial t}\left(\mathbf{B}\cdot\boldsymbol{\sigma}\right)=\boldsymbol{\sigma}\cdot\nabla\times\left(\mathbf{v}
\times\mathbf{B}\right)-\boldsymbol{\sigma}\cdot\nabla\times(\eta\jvec),
\]
or
\begin{align}
\frac{\partial}{\partial t}\left(\frac{g}{r}\right)&=\mathbf{B}\cdot\nabla\left(\mathbf{v}_{\perp}\cdot\boldsymbol{\sigma}\right)-
\nabla\cdot\left(\frac{g}{r}\mathbf{v}_{\perp}\right)+\mathbf{v\times B\cdot}\nabla\times\boldsymbol{\sigma}\nonumber \\
&-\boldsymbol{\sigma}\cdot\nabla\times(\eta\jvec)
.\label{eq:g-advection-compression}
\end{align}
For the axially symmetric case, with $n=0$, this equation takes the familiar
form $\partial B_{z}/\partial t=\mathbf{B}\cdot\nabla v_{z}-\nabla\cdot\left(B_{z}\mathbf{v}_{\perp}\right)+\nabla \cdot 
(\eta \nabla B_z)$. As we 
have noted, for $n\neq 0$ we have $g=B_z\left(1+O(\epsilon^2)\right)$. The first term on the right in 
Eq.~(\ref{eq:g-advection-compression}), like the 
first term in the $B_z$ equation, represents an analog of 
the effect of differential rotation, i.e.~variation of $\mathbf{v}_{\perp}\cdot\boldsymbol{\sigma}$ on $\chi$ surfaces.
The second term on the right in Eq.~(\ref{eq:g-advection-compression}) shows the advection and compression of the quantity 
$g/r$ by $\mathbf{v}_{\perp}$. 
The third term is proportional to $\nabla\times\boldsymbol{\sigma}$. 
Resistive relaxation of the profile of $g$ is represented by the last term on the right of 
Eq.~(\ref{eq:g-advection-compression}). 
In steady-state, the left hand side of Eq.~(\ref{eq:g-advection-compression}) is zero. 
The contribution of each term to the right hand side of Eq. (\ref{eq:g-advection-compression}) is shown in 
Fig~\ref{fig:g_evolution} 
for the nominal case of AFNB, which uses the $E\times B$ VBC. 
For this particular case, Fig.~\ref{fig:g_evolution} shows that the first and fourth terms 
are negligible while the second and third terms mostly balance 
each other except near the edge where the contribution from the resistive term and $\mathbf{B}\cdot\nabla\left(\mathbf{v}_{\perp}\cdot\boldsymbol{\sigma}\right)$ play a role.
All of the terms in Eq.~(\ref{eq:g-advection-compression}) are multiplied by $r^3$ in Fig.~\ref{fig:g_evolution} to regularize 
the plotted quantities at $r=0$. 
This balance between the compression/advection and the $\nabla\times\boldsymbol{\sigma}$ terms in the evolution of $g$ is prevalent in a wide variety of regimes that have been studied and thus, these two effects are mainly responsible for the observed structure of the $g$ surfaces. 
An interesting feature associated with path of the primary current is that the current has a concentration on the left 
side, the same side on which the helical flux is concentrated.
The actual path of the current  lines and the magnitude of $g$ are affected to a large degree by 
the velocity boundary conditions (Sec.~\ref{sec:VelBC}) and the resistivity profile 
(Sec.~\ref{sec:eta_profiles}). 

\subsection{Current line length\label{sec:Delta-z}}
A closely related issue of interest is the length of current streamlines entering through 
the electrodes at $r=r_w$ to address the issue of current leakage. 
In the periodic geometry employed in this paper, 
these streamlines must also exit at $r=r_w$, and there may be closed current lines, separated from the open lines either by a separatrix with an X-point or by
a last closed current surface that is tangent to the wall at $r=r_w$.
The current streamlines are found by integrating
\begin{equation}
\frac{d\mathbf{x}}{d\tau}=\mathbf{j},\label{eq:Current-streamlines}
\end{equation}
where $\tau$ is a parameter related to length along the current streamlines, $d\tau=dl/|\jvec|$. 
These current streamlines are the analogs of the magnetic field lines found by integrating $d\mathbf{x}/d\tau=\mathbf{B}$,
and the current lines lie on $g=\mbox{const.}$~surfaces just like the magnetic field lines lie on 
$\chi=\mbox{const.}$~surfaces. 

Of particular interest is the current line length or the axial displacement of the current streamlines defined as
\begin{equation}
\Delta z=\ointop\frac{j_{z}du}{j_{\perp}},\label{eq:DeltaZ}
\end{equation}
where $j_{\perp}=\jvec\cdot\nabla u$. 
The quantity $q_h$ of Sec.~\ref{sec:Pitch} is the analog of the quantity $\Delta z/L$ here, with magnetic field lines replacing current density streamlines, with one caveat: In $q_h$ (or $\bar{\iota}_h$), the field lines are integrated over the whole $\chi=\text{const.}$ surfaces whereas in Eq.~(\ref{eq:DeltaZ}) the current density lines are integrated from the first point at $(r,z)=(r_w, z_0)$ to the second at $(r,z)=(r_w, z_{final})$, yielding $\Delta z=z_{final}-z_0$.

We initialize a set of points at $r=r_w$ at a particular axial position $z=z_0$ over values of $\theta$ for $j_r<0$, \textit{i.e.}~where current enters the system. 
An important point is that these seed points are distributed uniformly in $g$, i.e.~with 
$\Delta g= g_{max}-g_{min}\approx (\partial g/\partial \theta)\Delta \theta \mbox{= const.}$, rather than uniformly in $\theta$, 
so that the area subtended by the angle between any two adjacent 
streamlines carries the same amount of current entering the plasma $\Delta I_r \propto \Delta g$. 
A current line length $\Delta z$ that exceeds the periodicity length $L=2\pi R$ suggests, in periodic geometry, a shunting of the primary current to one of the secondary electrodes at $z=0$ and $z=L$. Current lines 
with $|\Delta z|/L < 1$, on the other hand, suggest current that flows from one primary electrode to the other\cite{FootNote3}. 

The orbits of $\jvec$ that emerge from the seed points with $j_r<0$ at $r=r_w$ and evolve according to 
Eq.~(\ref{eq:Current-streamlines}) are plotted in
Fig.~\ref{fig:J_streams}a and compared against the surfaces of the helical field $g$ displayed in Fig.~\ref{fig:J_streams}b 
for a strongly-driven case ($\phi_0=0.6$) with $E\times B$ VBC. 
The orbits match well with the $g$ surfaces in the region of open $g$ surfaces, as they should. 
These orbits originating at $r=r_w$ cannot, by construction, trace out the 
regions of closed $g$ surfaces and hence the reason for the appearance of a 
large void 
region in Fig.~\ref{fig:J_streams}a. The closed current lines in Fig.~\ref{fig:J_streams}b are separated from the 
open current 
lines attached to the wall
by a last closed surface that is tangent to the wall, at the right ($\theta=0$). 
The number of contours within the closed surfaces make up a significant portion of the number of contours that connect across the wall. 
The enclosed current lies strictly in the $\lambda>0$ region, and thus the secondary current is likely greater 
than the total plasma current of the device $I_z$, which is subject to the cancellation between the oppositely flowing channels.

\begin{figure}
\centering
\includegraphics[width=0.48\textwidth]{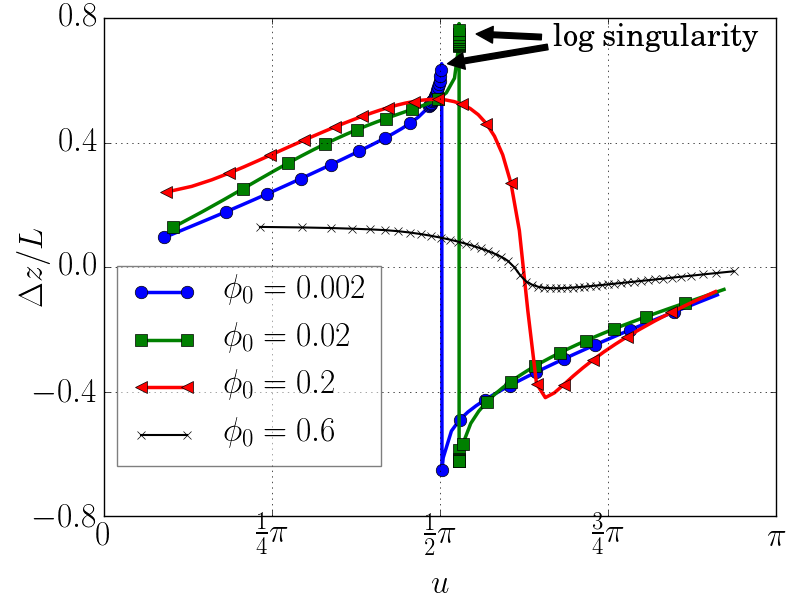}
\caption{\label{fig:DeltaZ_vs_Uinit} The axial displacement of the current density 
streamline orbits, $\Delta z=z_{final}-z_0$ relative to
the length of the cylinder $L$, as a function of the initial value of the helical coordinate $u$ for the nominal case of AFNB 
(dashed-dotted line) as well as three additional cases representing cases with weak drive ($\phi_0=0.002$), 
moderate drive ($\phi_0=0.02$), 
and strong 
drive ($\phi_0=0.6$), respectively.}
\end{figure}

Figure \ref{fig:DeltaZ_vs_Uinit} shows $\Delta z/L$ as a function of the initial value of the helical coordinate $u$. (The values of $\theta$ evident in 
Fig.~\ref{fig:J_streams} and $u$
in Fig.~\ref{fig:DeltaZ_vs_Uinit} correspond to $z=0$, so that, for $m=1$, $u$ equals $\theta$.) Results are shown for 
the nominal case of AFNB ($\phi_0=0.2$, red) as well as 
three additional cases with various $\phi_0$, all with $E\times B$ VBC. 
These cases have weak drive ($\phi_0=0.002$), 
moderate drive ($\phi_0=0.02$), and strong drive ($\phi_0=0.6$, as in Fig.~\ref{fig:J_streams}), respectively. 
These three cases were the subject of Section IIIC of AFNB and featured in Fig.~4 of that publication. These results show 
$\Delta z/L < 1$ and interestingly indicate 
a sharp drop in $\Delta z$ between $\phi_0=0.2$ and $\phi_0=0.6$. This appears to be related to the fact that over this range in $\phi_0$ the net current $I_z$ peaks and decreases by about $40$\%, as shown in Fig.~3 of AFNB, while $j_{\perp}$ and hence $\Delta g$ increase five-fold. 
The two cases with lower $\phi_0$ are nearly antisymmetric about the center, consistent with the fact that the axial current density (and $\lambda$) show two equal and opposite flows in the weak-to-moderate drive regime, 
similar to the early transient results shown in Fig.\ref{fig:transient}. 
The negative $\Delta z/L$ values represent the current streamlines that are mainly in the $\lambda<0$ ($j_z<0$ region). 

\begin{figure}
\centering
\includegraphics[width=0.48\textwidth]{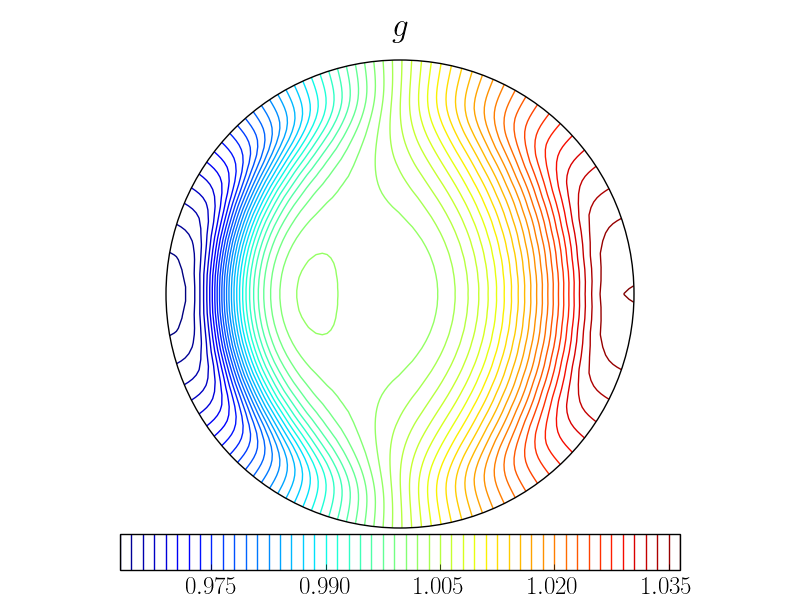}
\caption{\label{fig:g_phi002} The helical field $g$ surfaces for a weakly driven case ($\phi_0=0.02$ with $E\times B$ 
velocity BC) show a separatrix with two X-points, on the same $g$ surface because of up-down symmetry. The current 
streamline trace 
associated with this case is the green trace with squares in Fig.~\ref{fig:DeltaZ_vs_Uinit}, which shows a 
logarithmic singularity 
due to the X-points.}
\end{figure}

The two weak-drive cases also exhibit a singularity, which is related to the separatrix in the $g$ surfaces. 
This separatrix is visible in Fig.~\ref{fig:g_phi002}, for $\phi_0=0.02$.
The separatrix shown here has two X-points, which lead to a logarithmic singularity in $\Delta z$. Orbits near such a
separatrix have been shown to lead to logarithmic singularities\cite{LauFinn,LauFinn2}. 
The two X-points lie on a common separatrix
because of up-down symmetry; this symmetry has no effect on the logarithmic singularity.
Because of these singularities, we have $|\Delta z/L|>1$ for current streamlines that pass very close to the X-points, but the
range $\Delta \theta$ (or, more relevantly the range in current $\Delta g$) with $|\Delta z/L|>1$ is very small, and
we have $|\Delta z/L|<1$ for almost all of the streamlines.
No singularity appears for the nominal case, $\phi_0=0.2$ or for $\phi_0=0.6$; these cases appear to have tangencies
rather than separatrices in the $g$ contours, as seen in Fig.~\ref{fig:J_streams}.


\section{Sensitivity Studies\label{sec:Sensitivity}}
The focus of this section is a complete analysis of the sensitivity of the results to (a) the 
velocity boundary conditions, (b) the radial resistivity profile, and (c) the primary electrode shape/width. 
AFNB provided a brief overview of (a) and (b), which are revisited in more detail here. 
The subsequent sections will show that the current streamlines from the primary strongly depend 
on (a), moderately on (b), and not significantly on (c). 
The existence of pure secondary current flow (isolated from the helical electrodes) also depends 
strongly on (a) while (b) and (c) yield no qualitative changes.

\subsection{Velocity boundary conditions\label{sec:VelBC}}
\begin{SCfigure*}
\centering
\includegraphics[width=0.75\textwidth]{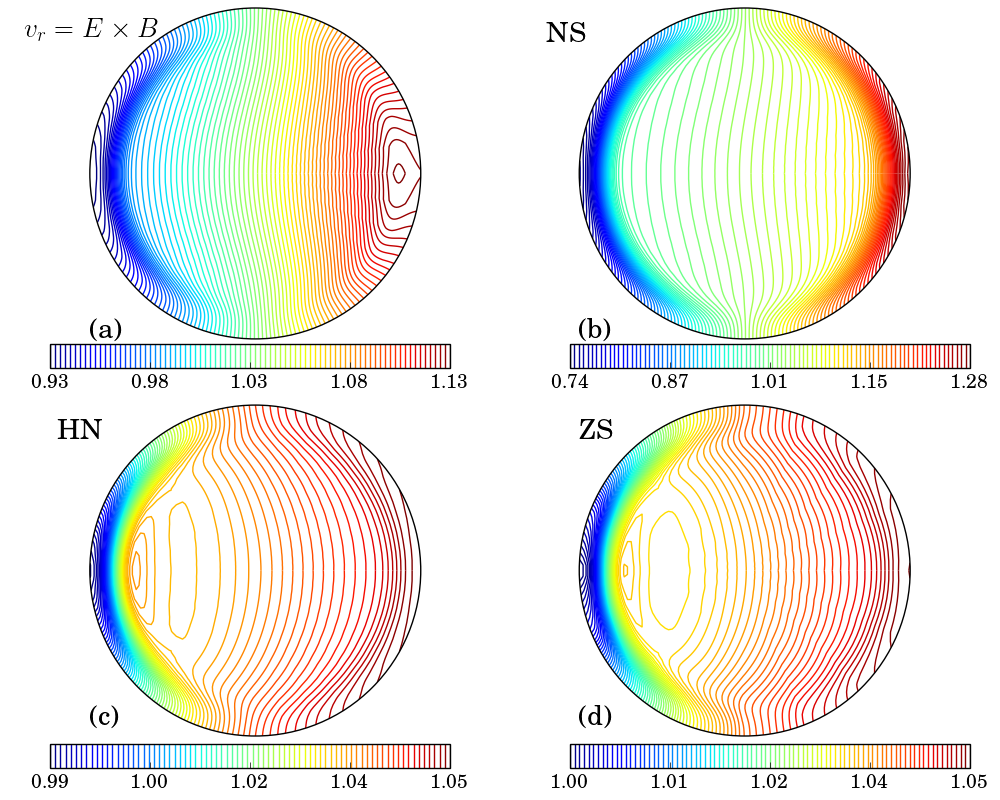}
 \caption{\label{fig:g-contours-4-cases} The contours of the helical field $g$ for the following four velocity boundary conditions (all with $\phi_0=0.2$): (a) $E \times B$ on the $(1,1)$ component of $v_r$ (the nominal case of AFNB); 
 (b) no-slip (NS); with (c) Neumann (HN) and (d) zero-stress (ZS). Case (a) has a tangency near
 $r=1,\theta=0$ with very little current on the closed lines; Case (b) has only open current lines; 
 Cases (c) and (d) have separatrices connected to $r=1$ for $\theta\approx2$ radians, but with a relatively small amount
 of closed current.}
\end{SCfigure*}

\subsubsection{Surfaces of the helical field $g$}
In this subsection we report on the sensitivity of the steady-state properties of the helical device of AFNB to changes in the 
velocity boundary conditions. 
A preliminary study of this sensitivity was performed in AFNB. There, the conditions used on the nominal case were: 
$E\times B$ boundary conditions on the $(1,1)$ Fourier components of $v_r$, no-slip BC 
on the remaining Fourier components of $v_r$ and on all components of $v_{\theta}$ and $v_z$. 
The results were compared with no-slip BC (NS), $\mathbf{v}(1)=0$ and HN boundary conditions, 
$\partial \mathbf{v}(1)/\partial r=0$.
The results discussed there indicated that while most quantities remained unchanged,
the helical field $g(r,u)$ varied considerably, in some cases exhibiting a set of closed current streamlines not connected to 
the wall at $r=r_w$ and 
in other cases showing no closed $g$ surfaces at all, depending on the boundary conditions. 
Furthermore, it was found that these closed current surfaces may be separated by a 
separatrix with an X-point or may have their outermost 
closed surface tangent to the boundary.

The surfaces of the helical field $g$ are shown in Fig.~\ref{fig:g-contours-4-cases} for the three aforementioned choices 
of the VBC as well as a fourth option: zero-stress on the tangential flow (ZS) with $E\times B$ drift imposed on the $(1,1)$ component $v_r(1)$.
All of these cases were driven with a helical potential of magnitude $\phi_0=0.2$. 
Results show that the current proportional to $\nabla g \times \boldsymbol{\sigma}$  
on closed $g$ surfaces is very small for the nominal $E\times B$ case (Fig.~\ref{fig:g-contours-4-cases}a). 
The no-slip (NS) boundary conditions display no closed surfaces for this (Fig.~\ref{fig:g-contours-4-cases}b) or 
any value of $\phi_0$. 
The HN and ZS boundary conditions feature an area of closed $g$ surfaces, located at a 
different position than for the $E\times B$ case, and separated from the open $g$ surfaces by a separatrix. This separatrix
has two X-points, on the same $g$ surface, as discussed.
The enclosed current within the closed $g$ surfaces grows as $\phi_0$ increases for all choices of VBC except NS. 

An important factor is the difference between the net primary current $\Delta g = g_{max}-g_{min}$ for the different 
VBC. From Fig.~\ref{fig:g-contours-4-cases}, we have $\Delta g=0.20,0.54,0.06$, and $0.05$ for the four cases.
The most striking effect is that the net current is much larger for the no-slip case. This appears to be
due to the fact that if the plasma flow is constrained to be zero at the wall, the applied voltage at $r=r_w$ 
can only drive current there. In the other cases, in which the plasma is free to move, much of the applied voltage 
causes $E\times B$ motion (with a correction proportional to $\mathbf{j}\times \mathbf{B}$, as
discussed in the Appendix.) As we show below in Sec.~\ref{sec:streams} these differences have a direct bearing 
on how far the current streamlines travel in the axial direction.

Results in AFNB and studied in more detail in Sec.~\ref{sec:VelBC}
show that the area of closed current lines disappears when no-slip (homogeneous Dirichlet) boundary conditions are applied. 
Further, results in Sec.~\ref{sec:VelBC} show that a large area of closed current lines can be present when 
homogeneous Neumann boundary conditions are employed.

Similar results to those shown in Fig.~\ref{fig:g_evolution} were found with the last three sets of VBC, showing some 
small differences. However, the main qualitative conclusion, namely that the second and third terms mostly balance with 
a small correction from the resistive term at the edge, holds for all four VBC.
 
Because of the stark difference in the velocity profiles, especially in $v_r$ and $v_{\th}$
the input power is 
predominantly dissipated \textit{via} two different channels for NS \textit{vs} HN or ZS VBC. 
For the latter two conditions, $80-90\%$ of the input power is Ohmically dissipated while for no-slip--because of the strongly non-uniform velocity profiles--the viscosity dissipates $\sim 65$\% of the input power in steady-state, similar to the steady-state power partition for the 
nominal case discussed in Sec.~\ref{sec:Transient} and depicted in Fig.~\ref{fig:power_vs_time}


\subsubsection{Current density streamlines\label{sec:streams}}
\begin{figure}
\centering
\includegraphics[width=0.48\textwidth]{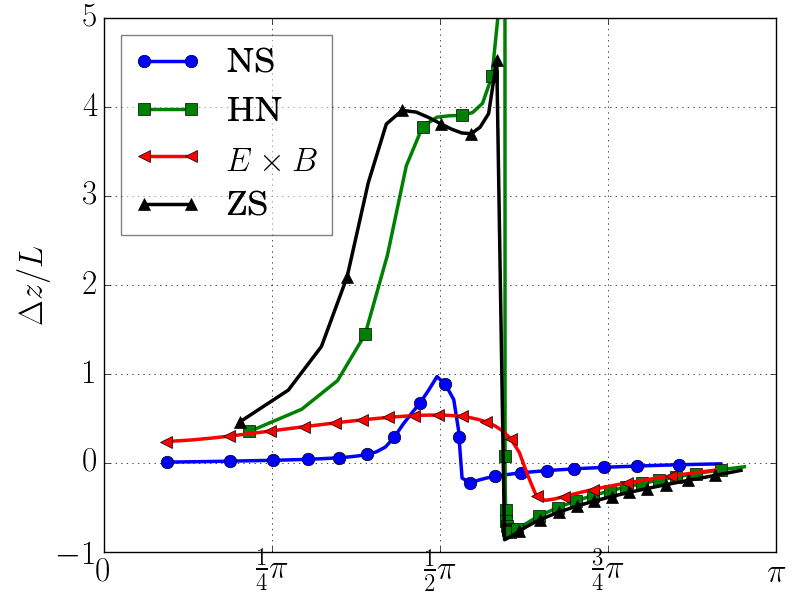}
   \caption{\label{fig:DeltaZ_vs_Uinit_VelBC} The axial displacement of the current density streamline orbits, 
   $\Delta z=z_{final}-z_0$ 
   relative to
   the length of the cylinder $L$, as a function of the initial value of the helical coordinate $u$. Results shown 
   for four cases that utilize 
   $E\times B$ condition on the $(1,1)$ component of $v_r$ (blue circles), no-slip condition (green squares), and homogeneous Neumann condition (red triangles), and zero-stress condition respectively, all with  $\phi_0=0.2$.}
\end{figure}
In this section we investigate the changes in the current density lines caused by the changes in velocity boundary conditions
studied in the last section. 
We have noted that there is a significant difference in $\Delta g$ for each choice of velocity BC in 
Fig.~\ref{fig:g-contours-4-cases}, with the no-slip case showing 
the largest values of $\Delta g$, with the cases (HN and ZS) yielding the smallest values of
$\Delta g$, while $j_z$ and thereby the total secondary current $I_z$ are similar in all four cases.

Figure \ref{fig:DeltaZ_vs_Uinit_VelBC} shows $\Delta z/L$ as a function of $u$ (vs.~$\theta$ for $z=0$, $m=1$), 
for the nominal case ($\phi_0=0.2$) of AFNB and 
three additional cases that employ the three remaining aforementioned velocity boundary conditions: NS, HN, 
and ZS, all driven with $\phi_0=0.2$. 
The axial excursion $\Delta z$ of the current streamlines is significantly larger for HN and ZS than NS. 
The NS case mostly has $|\Delta z|/L<<1$ as a consequence of zero plasma motion at the boundary, which leads to a large $j_r(r_w)$ as well as a large $j_{\perp}$ and $\Delta g$ and therefore a smaller $j_z/j_{\perp}$ and $\Delta z$ as indicated by Eq.~(\ref{eq:DeltaZ}).
HN and ZS have nearly identical behavior with $\Delta z/L>1$ over an appreciable range, $0.35\pi 
\lesssim \theta \lesssim 0.7\pi$, 
showing a 
significant fraction of current from $r=r_w$ traveling far enough in $z$ to represent shunting, \textit{i.e.} current 
flowing directly from the primary to the secondary.
Both ZS and HN conditions exhibit a logarithmic singularity in $\Delta z/L$ around $u=0.6\pi$, due again to the presence of a separatrix in $g$ surfaces, but again the amount of current near the singularity is small.

\begin{figure}
\centering
\includegraphics[width=0.48\textwidth]{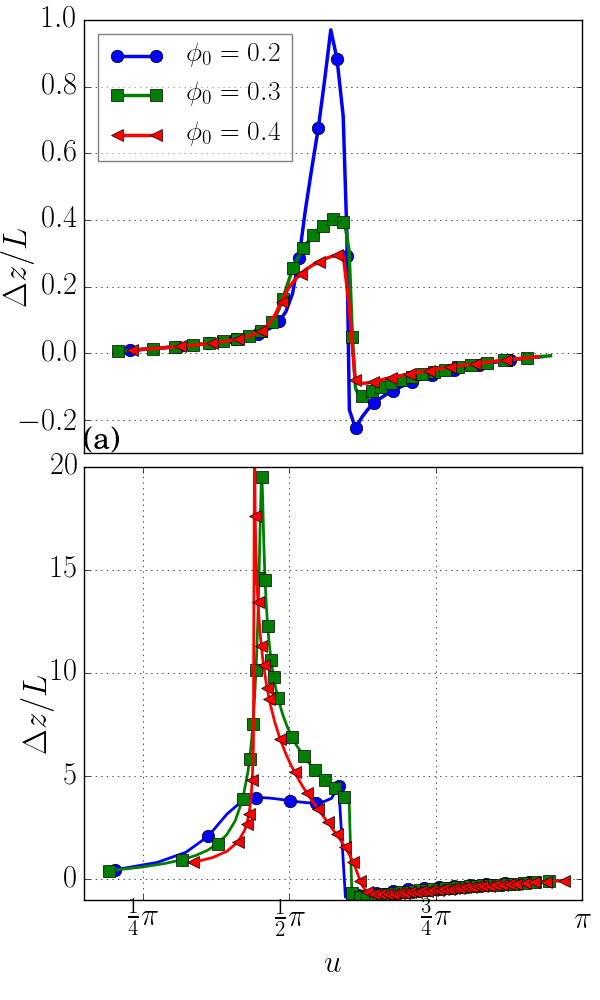}
\caption{\label{fig:HD_NS_deltaz} The axial displacement of the current density streamline orbits, 
$\Delta z/L$, as a function of the initial value of the helical coordinate $u$ 
for various values of $\phi_0$ in the strong drive regime. Results are shown for (a) no-slip (NS) and (b) 
zero-stress (ZS) velocity boundary conditions. }
\end{figure}

The trends for increasing $\phi_0$ in the strong drive regime for NS and ZS cases are shown in Figs.~\ref{fig:HD_NS_deltaz}a and b, respectively. 
For NS, $\Delta z/L<1$ and noticeably and monotonically decreases as the drive strength increases. 
However, this trend does not hold for ZS (and similarly for HN) as indicated by Fig.~\ref{fig:HD_NS_deltaz}b. 
For $\phi_0 = 0.2$ we see that $ \Delta z/L \approx 4$ over a region. 
Near $u=0.62 \pi$ there is a logarithmic singularity due to a separatrix with an X-point, as seen in Fig.~9d.  
At this same value of $u$ there is also a jump down to negative $\Delta z$ values. 
This corresponds to the $\lambda<0$ ($j_z<0$) region where the majority of the current streamlines reside. 
The current streamlines are usually much shorter in this region because of the very dense packing of the $g$ surfaces (\textit{c.f.} the left side of all of the $g$ figures shown in this paper), indicating a very large $\Delta g\propto j_{\perp}$, and hence a small and negative $j_z/|j_{\perp}|$. 
There is evidence of strong shunting for $\phi_0 =0.3$ and $0.4$, with $\Delta z/L \gg 1$ over a significant region. 
The discontinuity in $\Delta z$ at $u \approx 0.62 \pi$ due to the separatrix is still apparent for $\phi_0=0.3$, but is absent for $\phi_0=0.4$, indicating that this separatrix region disappears for $\phi_0>0.3$ as corroborated by the $g$ contours for each case (not shown). 
In addition, each of the $\phi_0 = 0.3$ and $\phi_0 = 0.4$ traces exhibits a strong spike just below $u = \pi/2$. 
For the latter a closer look shows that the spike corresponds to a singularity associated with the separatrix of a closed $g$ region formed by a bifurcation just below $\phi_0 =0.4$. 
For the former, $\phi_0 = 0.3$ is just below the bifurcation point and the apparent singularity represents a small region of very long current streamlines with $\Delta z$ smooth.
Also note that for ZS driven with $\phi_0 = 0.3$ and $0.4$, the separation of the initial current streamline points at $r=r_w$, having $\Delta g$ constant, shrink in range because the volume of closed $g$ surfaces grows as $\phi_0$ increases. 
This feature is absent for the concentrated electrode conﬁgurations, where $j_r$ is prescribed to be emitted from only a fraction of the wall corresponding to the physical electrodes.

\subsection{The effect of the resistivity profile\label{sec:eta_profiles}}
\begin{figure*}
  \centering
  \includegraphics[width=0.9\textwidth]{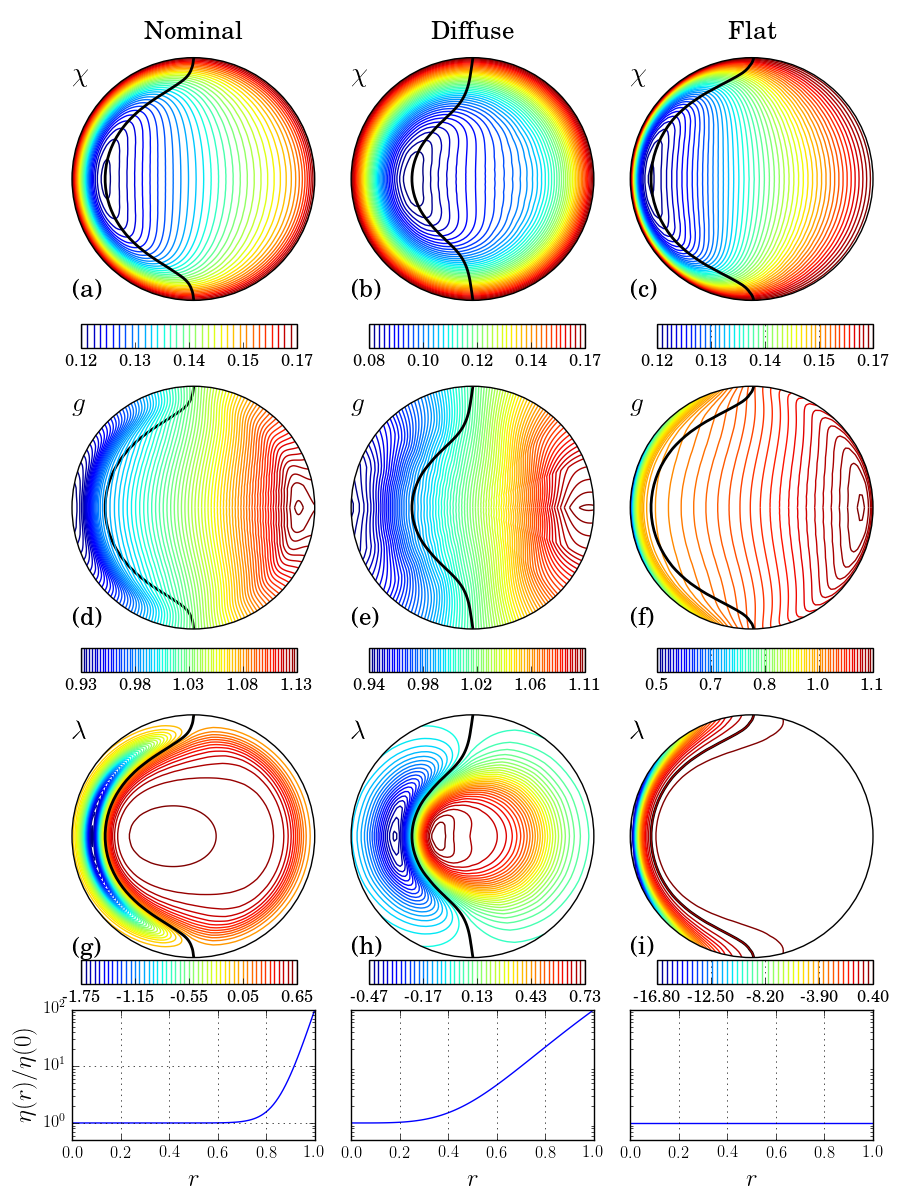}
  \caption{\label{fig:eta_profs} The contours of helical flux $\chi$ (top row), helical field $g$ (2nd row), and parallel current density $\lambda$ (3rd row) are shown for AFNB's nominal (hollow) resistivity profile, the diffuse profile, and (c) the flat profile. The bottom row shows the normalized resistivity profile for each case on a logarithmic scale. The black curves correspond to $\lambda=0$. Note the concentration of $g$ surfaces
  to the left of $\lambda=0$.}
\end{figure*}
\begin{SCfigure*}
  \centering
  \includegraphics[width=0.8\textwidth]{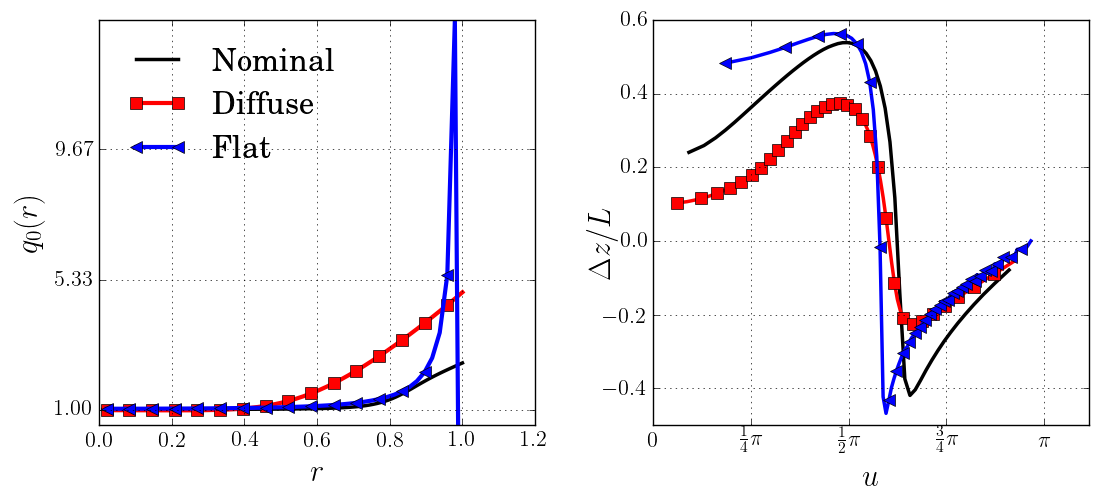}
  \caption{\label{fig:q0_DeltaZ_eta_profs} Mean safety factor profile $q_0(r)$ and $\Delta z(u)/L$ with $E\times B$ 
  velocity BC shown for the
  nominal (hollow) profile, the diffuse profile, and the flat resistivity profile. }
\end{SCfigure*}
Simulations with two additional resistivity profiles were run with three of the aforementioned velocity boundary conditions 
to study the influence of the profile $\eta(r)$ on the characteristics established in the earlier sections and in AFNB. 
We compare the nominal or hollow profile used in the previous sections and in AFNB with $p=16$, $\eta(r_w)/\eta(0)=100$ in 
Eq.~(\ref{Eq:resistivityprofile}) with a diffuse profile
($p=4$, $\eta(r_w)/\eta(0)=100$) and with a flat profile ($p=0$, $\eta(r_w)/\eta(0)=1$.)
A comparison of the first and third of these profiles was made in Section IVB of AFNB. 

Results for the three resistivity profiles with the $E\times B$ velocity boundary conditions are shown in 
Fig.~\ref{fig:eta_profs} for $\phi_0=0.2$. 
The bottom row of Fig.~\ref{fig:eta_profs} shows the normalized resistivity profiles on a logarithmic scale. 
The value of the edge resistivity $\eta(1)$ is the same for both the hollow and diffuse cases. 
The surfaces of the helical flux $\chi$ show a similar level of shift and distortion for the nominal 
(hollow) and flat resistivity profile 
(Fig.~\ref{fig:eta_profs}a and c). 
All three cases produce `D'-shaped and elongated $\chi$ surfaces near the O-point. The
flux surface distortion is strongest for the 
flat case, with $(r_O=0.83)$. This distortion is moderate for the hollow case $r_O=0.7$ and weakest for the diffuse case, $(r_O=0.5)$. 
These results are plausible because the position of the O-point
is determined by where the advection of flux is stopped by resistive diffusion. 

The helical field $g$ contours are displayed in the second row (from the top) of Fig.~\ref{fig:eta_profs}. The effect of the 
resistivity profile is much more profound on the $g$ surfaces than on the helical flux. 
In particular, compared to the hollow case of Fig.~\ref{fig:eta_profs}d,
the diffuse profile case of Fig.~\ref{fig:eta_profs}e
relaxes the concentration of the primary current path on the left. The flat profile shown in Fig.~\ref{fig:eta_profs}f
forces the open current to flow mostly on the left side of the radial wall, with a significantly larger volume of closed $g$ surfaces compared to the other two profiles. The range in $g$ is also enhanced for the flat resistivity case since the edge resistivity is now much lower, resulting in an 
increased amount of net current ($\propto \Delta g$ entering and exiting the domain. 
It is interesting to note that in all three cases the most concentrated
area of $g$ contours is in the $\lambda<0$ region.

The parallel current $\lambda$ for the diffuse case (Fig.~\ref{fig:eta_profs}h) shows more evenly 
balanced channels of positive and negative current density
than the hollow profile case  (Fig.~\ref{fig:eta_profs}g),
consistent with the more modest displacement of the $O-$point. On the other hand, the values of 
$\lambda$ become far more negative in the $\lambda<0$ region of the flat resistivity case shown in Fig.~\ref{fig:eta_profs}i. 
These results are consistent with the average of $\eta\lambda B^2$ along the O-line and on the flux surfaces $\chi=$const., as discussed in AFNB.
The flat resistivity profile (Fig.~\ref{fig:eta_profs}i) flattens the vales of $\lambda$ in the
$\lambda>0$ region (Fig.~\ref{fig:eta_profs}i), with a maximum value about $28$\% lower than for the hollow profile.

The profile of $q_0(r)$ and a plot of $\Delta z(u)/L$ are shown in Fig.~\ref{fig:q0_DeltaZ_eta_profs}. 
The nominal (hollow) and flat resistivity profiles both produce a very flat profile of $q_0$ throughout most of the plasma.
For the flat resistivity case $q_0(r)$ goes to infinity near the wall because $B_{\th}^{(0,0)}$ goes to zero there.
That is, the net current is close to zero in the flat resistivity case; 
The $q_0(r)$ profiles for the hollow and flat cases were also shown in Fig.~1 of AFNB. 
The diffuse resistivity profile results in a more diffuse and sheared $q_0$ profile, with $q_0(r)\approx 1$ for a 
somewhat smaller region; The overall larger resistivity for the diffuse case has results 
as if the drive strength were reduced, 
consistent with a more modestly displaced $O-$ point from the cylindrical axis. 
The current streamline length, $\Delta z/L$, has a weak dependence on the resistivity profile. 
The diffuse case has somewhat smaller values of $\Delta z/L$ due to the increased effective plasma resistivity, which reduces $I_z$(nearly halves it in this case) and causes a smaller $j_z/j_{\perp}$. 

Perhaps a more consistent way to measure the effect of resistivity profile on the characteristics is one based on equal volume-averaged resistivities $\int_0^{r_w} \eta(r)r dr/r_w^2$ for all 3 cases instead of one where $\eta(0)$ is kept the same. 
This amounts to an $\eta(1)/\eta(0)\approx30$ for the diffuse case and a tripling of $\eta(0)$ for the flat case (the exact multiplication factor is 3.3). 
For the latter, the increase in $\eta(0)$ also implies reducing $S$ by a factor of 3.3 to $S=30$. 
As expected, modification of the diffuse profile according to this prescription yields both a greater $I_z$ and $r_0$, changes that simply amount to a slightly stronger helical drive, while increasing the resistivity 3.3-fold in the flat case results in a weaker helical drive. 

The velocity boundary conditions affect the dynamics in the same manner for each case regardless of the resistivity profile (not shown here). 

We conclude that while there are some quantitative differences in the results due to resistivity profiles, the major characteristics change very little qualitatively. 
From an application perspective, the cancellation in $I_z$ for example, is not a factor because it can be mitigated by slotting of the secondary electrodes, as discussed in AFNB. 

\subsection{Concentrated-electrode configuration\label{concentrate}}
In order to model the electrodes of the experimental device at Tibbar Plasma Technologies,
or any such device with physical electrodes, more accurately the helical drive is modified in this section  to 
simulate a pair of thin helical primary surfaces. 
The simplest implementation involves specifying the normal component of current density, $j_r(u)$ at the wall. 
The concentration of the surface $j_r$ requires a broad spectrum of Fourier harmonics, which must be truncated in the MHD simulations. 
The prescription specifying a single Fourier component of 
$j_r(r_w)$, $(m,n)=(1,1)$, has been described in AFNB, but with zero tangential electric field 
on the other components rather than specifying $j_r(r_w)=0$ for these components. 
For comparison, the previous prescription with $(m,n)=(1,1)$ is called a ``smeared electrode''.

Only harmonics of $j_r(r_w)$ with $m/n=1$ occur because of the single helicity application. Odd parity about $\theta=0$ (for $z=0$) implies only $\sin$ terms contribute. We also assume symmetry of $j_r(r_w)$ about $\theta=\pi/2$, so that only harmonics with odd $m$, namely $(m,n)=(1,1),(3,3),(5,5)\cdots$ occur.
Clearly, a narrower electrode requires a greater number of Fourier harmonics.
The baseline simulation assumes the width of each of the two electrodes spans $(\Delta \theta)_0=30$ degrees 
of the cylindrical boundary. 
Cases with $(\Delta \theta)_0=45^{\circ}$ and $(\Delta \theta)_0=20^{\circ}$ electrode span were also simulated to chart the 
sensitivity of defining characteristics to the electrode width. 
The angular span of the experimental electrodes is smaller, approximately $10-15^{\circ}$. 
Simulations have been run again with no-slip (NS), homogeneous Neumann (HN), and zero-stress (ZS) boundary 
conditions (with $E\times B$ VBC omitted for this study.)
An alternate formulation in terms of the electrostatic potential at $r=r_w$ has been developed, 
mainly for benchmarking with the NIMROD code, and will be described in a future publication.

\begin{SCfigure*}
\centering
  \includegraphics[width=0.75\textwidth]{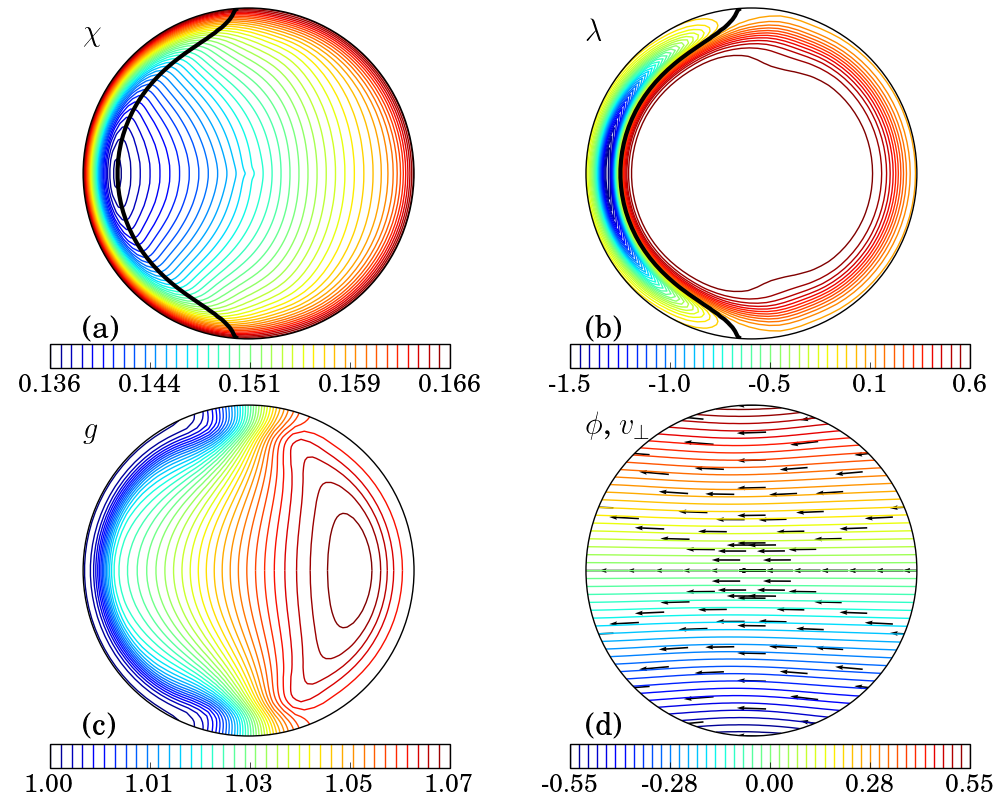}
 \caption{\label{fig:localElect1} Concentrated electrode case with zero-stress (ZS) velocity boundary conditions
 and $(\Delta \theta)_0=30^{\circ}$. Shown are the contours of (a) The helical flux $\chi$, (b) the parallel current density $\lambda$, 
 (c) the helical field $g$, 
 and (d) the electrostatic potential $\phi$ 
(with vectors of $\vperp$) for a strongly-driven simulation (with $||\vperp||/v_A=0.54$). 
The mean safety factor is $q_0(r)=1.02$ in the plasma interior. Note that, as required, no $g$ contours intersect the wall outside the electrode region defined by $(\Delta \theta)_0$ in (c).}
\end{SCfigure*}

The helical flux $\chi$, parallel current density $\lambda$, helical field $g$, and the electrostatic potential $\phi$ (with 
vectors of $\vperp$) are shown in 
Fig.~\ref{fig:localElect1} for a strongly-driven case with an electrode span of $(\Delta \theta)_0=30^{\circ}$ and NS VBC. 
11 odd Fourier harmonics were required to specify $j_r(r_w)$ for this case, which employed a total number of 22 Fourier harmonics in the actual simulation. 
De-aliasing of the quadratic nonlinearities result in the discrete number of modes stated here: \textit{e.g.} 22 for $(\Delta \theta)_0=30^{\circ}$ and 43 $(\Delta \theta)_0=20^{\circ}$. 
The drive strength is equivalent to a smeared-electrode configuration driven with $\phi_0=0.5-0.6$, based on the 
location of $O-$point ($r_O=0.79$) and magnitude of the bulk flow ($||\vperp||/v_A=0.54$). 
See Figs.~4e and for AFNB (run with $E\times B$ VBC).
The mean safety factor $q_0$ is flat over approximately 80\% of the radius with $q_0(r=0)=1.02$. 
Note that very few contours of $g$ intersect the wall ($r=r_w$) outside the electrode region determined by 
$(\Delta \theta)_0$ in this figure, as expected.
There appears to be a large volume containing closed $g$ surfaces (Fig.~\ref{fig:localElect1}c). 
The size of this volume is consistent with that observed in a smeared-electrode configuration that was run with ZS VBC and at nearly the same electrostatic drive strength.

 
Simulations with other values of electrode width $(\Delta \theta)_0$ show similar results.
Two additional simulations run with $(\Delta \theta)_0=45^{\circ}$ and $(\Delta \theta)_0=20^{\circ}$, employing ZS VBC (not shown here), that match $||\vperp||/v_A$ of the case displayed in Fig.~\ref{fig:localElect1} both produce a closed $g$ region of similar size to the one observed for $(\Delta \theta)_0=30^{\circ}$ illustrated in Fig.~\ref{fig:localElect1}c. 
In addition, very similar $\chi$ and $\lambda$ structures and identical values of $r_0$, $q_0(0)$ are obtained in both cases.  

Simulations performed with different VBC and the resistivity profiles of Sec.~\ref{sec:eta_profiles} (not shown) yield no significant qualitative differences.

In summary, the quantities shown in Fig.~\ref{fig:localElect1} all look very similar to their 
smeared electrode counterparts, display the same trends such as a very flat $q_0$ profile in the interior with 
$q_0(0)\rightarrow 1.0^+$ in the strong-drive ($||\vperp||/v_A\gtrsim 0.2$) regime as the $j_r(r_w)$ is increased.
Thus, the characteristics of this helical state are robust to the electrode width. 
 
\section{Summary\label{sec:Summary}}

In this paper we have extended the findings of AFNB on the electrostatically driven helical plasma, specifically studying 
the variation of quantities that are of potential importance to the DC electrical transformer application discussed in AFNB. 

We have studied the transient leading to the final steady state, focusing on the magnetic helicity and on the channels that 
dissipate the input power during the transient. We conclude that both in the early 
and late phases of the transient the 
flux surface average 
$\langle \eta \lambda B^2 \rangle\simeq 0$, corresponding to zero net helicity injection rate. 
On the other hand, there is a middle stage in the transient in which the volume average of
$\eta \lambda B^2<0$, leading to a positive helicity rate. 
It is during this period that the helicity production rate peaks. 
There is no helicity injection from the 
boundary since $B_n=0$ there. However, the 
region with $\lambda<0$ generates helicity 
which is injected into the $\lambda>0$ region, where helicity is dissipated. 
For the nominal case of AFNB and this paper, with $E\times B$ VBC, the input power is dissipated mostly 
Ohmically during the transient and then, via viscous losses 
during the time-asymptotic state. 
This particular finding depends on the velocity boundary condition at the radial wall and homogeneous Neumann or zero-stress 
boundary conditions result in minimal viscous dissipation, so that the major losses are Ohmic.

We have studied effects related to tracing magnetic field lines and current density lines, specialized to $(m,n)=(1,1)$. For 
the former, we have computed the 
helical transform $\bar{\iota}_h(\chi)$, the change $\Delta u$ in $u=m\theta+kz$ over one period relative to $\Delta z$, the 
change in $z$ over one flux surface
labeled by $\chi$. This quantity is the analog 
of the rotational transform $\bar{\iota}\equiv\Delta \theta/\Delta z$ in toroidal geometry. 
In addition, we have $\bar{\iota}_h(\chi)=\bar{\iota}(\chi)-1$. 
This is related to the separation of the  helical behavior of the field lines into twist 
and writhe\cite{Bellan-textbook}, the latter represented by $n/m=1$ in this formula. 

Regarding current lines, we have developed two metrics: one relating to the secondary-to-secondary current, and the other to the 
possible shunting of the primary current. 
Metric (1) focuses on the helical field $g$ and terms in the steady-state Ohm's law that affect the evolution of the surfaces of $g$, Eq.~(\ref{eq:g-advection-compression}).
Closed contours of $g$, disconnected from the electrodes by an X-line or by a tangency at $r=r_w$,
indicate the possibility of pure secondary-to-secondary current. 
Metric (II) integrates along the primary current streamlines to measure their axial length (displacement) $\Delta z$. 
Streamlines with $\Delta z$ exceeding the periodicity length $L=2 \pi R$ suggest the undesirable direct flow of electrical 
current between the primary and secondary electrodes, \textit{i.e.} ``shunted'' current. 

A major topic of this paper has been an analysis of the sensitivity of the application to (a) the 
velocity boundary conditions, (b) the resistivity profile, and (c) the electrode width. 
AFNB provided a brief overview of (a) and (b). 
This sensitivity to these factors is gauged in terms of the two concepts introduced in Sec.~\ref{sec:CD-streamlines} 
and again in the previous paragraph.  
We found that a secondary current, indicated by the occurrence of closed helical field $g$ surfaces with a tangency or 
a separatrix can occur, depending on the type of velocity boundary condition employed.  
The $E\times B$ drift imposed on the $(1,1)$ component of the radial velocity results in a closed $g$ volume with a 
tangency while homogeneous Neumann (HN) and zero-stress (ZS) conditions on the 
boundary velocity result in closed $g$ surfaces with 
a separatrix. 
The volume of the closed $g$ region, proportional to the magnitude of the secondary current, increases as the electrostatic drive is turned up. 
No closed $g$ are observed for no-slip (NS) boundary conditions. 
We have also traced the current density streamlines. The open streamlines, connected to the wall, 
indeed lie on the open constant $g$ surfaces.
For $E\times B$ or no-slip boundary conditions, $|\Delta z|/L<1$ and for NS $|\Delta z|/L\ll1$. However, for HN and ZS boundary conditions the current associated with $g$ is small, so that $|\Delta z|$ can exceed $L$. 
In all cases, the presence of a separatrix in $g$ surfaces leads to
a logarithmic singularity in $\Delta z$, although current lines near this singularity do not represent a significant amount of 
current. 
Cases with closed $g$ surfaces separated by a tangency do not show noticeable features in $\Delta z$ due to this tangency.

Sensitivity to the resistivity profile was charted by employing a diffuse resistivity profile with 
$\eta(r_w)/\eta(0)=100$ in addition to the hollow (nominal) and flat profiles of AFNB. 
The flat profile leads to increased distortion of helical flux relative to that of the hollow resistivity profile, a larger radial displacement of the $O-$point (bigger $r_O$), 
a flatter $q_0(r)$, and  $I_z\simeq0$. 
The diffuse profile leads to only modest shift/distortion of flux surfaces relative to the hollow profile, with a smaller value of $r_0$, a smaller $|\Delta z|/ L$, and a larger gradient in $q_0(r)$ except near $r=0$. 
The smaller $|\Delta z|/ L$ is simply due to the increased effective resistivity, which lowers the plasma current density $j_z$ and thus, tips the ratio $j_z/j_{\perp}$ in favor of $j_{\perp}$. 
Holding the average resistivity constant between the three cases leads to similar conclusions. 

A third and final sensitivity study focused on the departure from the sinusoidal (``smeared'') 
electrostatic drive employed in AFNB 
(and in earlier sections in this paper) to one that models concentrated primary electrodes, 
electrodes which are localized in both $\th$ and in $z$.
Nearly all of the previously observed qualitative behavior that pertains to the smeared electrode configuration, 
including the response to the velocity boundary conditions and resistivity profiles, remains unchanged for the simulations run with a concentrated-electrode configuration.

Future work will investigate steady-state solutions of the helical drive with harmonics $(m,n)$ other than $(1,1)$ and the dependence of the above characteristics 
on the aspect ratio for the periodic cylinder. 
A subsequent publication will feature imposing a non-zero normal magnetic field $B_r(r_w)$ at the wall to implement electrostatic helicity injection and documenting the efficiency of the DC transformer device with and without $B_r(r_w)$. 
This will be followed by studying the same electrostatic helical drive in a cylinder of finite length where the surrounding walls could be perfectly conducting or resistive. 

\section*{Acknowledgements}
We thank 
Aaron McEvoy, Juan Fernandez, William Gibson, Keith Moser, and Liviu Popa-Simil for their input and valuable 
discussions. 
This research was supported by funding from the ARPA-E agency of the Department of Energy under Grant No. DE-AR000067


\section*{Appendix: Contributions to the perpendicular plasma velocity; zero-stress boundary conditions} 

From Ohm's law, Eq.~(\ref{eq:Ohm}), the perpendicular velocity equals
\[
\mathbf{v}_{\perp}=\frac{\mathbf{E}\times\mathbf{B}}{B^{2}}-\eta\frac{\mathbf{j}\times\mathbf{B}}{B^{2}}
\]
To compare the magnitude of the first term, the $E\times B$ drift,
with the correction term, we will simplify by assuming either that
the Lorentz force is balanced by inertia or by viscous stresses. Assuming
the first case and also $|v_{||}|\ll|v_{\perp}|$ (as observed), 
we find $\mathbf{j}\times\mathbf{B}\sim\rho_{0}v_{\perp}^{2}/r_{w}$,
so we have
\[
\frac{\mathbf{E}\times\mathbf{B}}{B^{2}}\,\,::\,\,\eta\frac{\mathbf{j}\times\mathbf{B}}{B^{2}},
\]
\[
v_{\perp}\,\,::\,\,\frac{\eta\rho_{0}v_{\perp}^{2}}{B^{2}r_{w}}
\]
or
\begin{equation}
1\,\,::\,\,\frac{v}{v_{A}}\frac{1}{S}.\label{eq:comparison-inertia}
\end{equation}
Thus, the resistive correction to the $E\times B$ drift is small
unless both $v/v_{A}\simeq 1$ and $S\lesssim1$. For most
of the parameters we use, this correction is small, and we expect
the perpendicular velocity to be well approximated by the $E\times B$
drift. 
But since our interest is in a correction at $r=r_w=1$, the Lundquist number $S$ that appears in Eq. (\ref{eq:comparison-inertia}) must be based on the edge resistivity, $\eta(1)$, which implies $S(1)=1$ for the nominal parameters: $S=100$ and $\eta(1)/\eta(0)=100$. 
Thus, the correction term reduces to $v/v_{A}$, which is small except for the most strongly driven cases.

For the other case, in which the Lorentz force is balanced by viscous
stresses $\mathbf{j}\times\mathbf{B}\sim\mu v_{\perp}/r_{w}^{2}$,
the comparison is
\[
1\,\,::\,\,\frac{\tau_{A}}{\tau_{v}}\frac{\tau_{A}}{\tau_{r}},
\]
\[
1\,\,::\,\,\frac{1}{S}\frac{1}{Re}.
\]
Once again, applying this expression to the plasma edge where $S=1$ and $Re=10$ (unchanged) results in a correction factor of $0.1$. 
If the Lorentz force is balanced by both inertia and viscous stresses, a reasonable approximation is that the correction is the larger of the corrections for the two cases. 
We conclude that the resistive correction to the $E\times B$ drift is small, unless $v/v_A\sim 1$ or $Re\lesssim 1$. 
For either case, the resistive correction is smaller for higher edge $S$.

The conventional zero-stress boundary conditions on the tangential components of the velocity
are obtained from the symmetrized stress
tensor $\Pi_s=\mu\left(\nabla 
\mathbf{v}+(\nabla \mathbf{v})^T \right)$ discussed in Sec.~\ref{sec:Model}. 
Zero-stress boundary conditions at $r=r_w$ corresponding 
to this symmetric stress tensor are of the form\cite{LandauLifshitz}
\begin{equation}
     \mu\left(\frac{im}{r}v_r+\frac{\partial v_{\theta}}{\partial r}-\frac{v_{\theta}}{r}\right)=0\label{eq:no-stress-theta}
\end{equation}
and
\begin{equation}
     \mu\left(\frac{\partial v_z}{\partial r}+ikv_r\right)=0.\label{eq:no-stress-z}
\end{equation}
Using the non-symmetrized stress tensor discussed in Sec.~\ref{sec:Model}, namely
$\Pi=\mu\nabla \mathbf{v} $, which leads to the viscous operator from Eq.~(\ref{eq:Momentum}) used in the DEBS code,
we find zero-stress boundary conditions of the form
\begin{equation}
     \mu\left(\frac{\partial v_{\theta}}{\partial r}\right)=0 \mbox{ ,   }
     \mu\left(\frac{\partial v_z}{\partial r}\right)=0.\label{eq:zero-stress-z}
\end{equation}


\begin{thebibliography}{10}
\bibitem{FootNote1}
Interestingly, at the times for which $q_0(0)$ is significantly less than
  unity, no islands in the $\chi$ surfaces are evident.
\bibitem{FootNote2}
Alternatively, we can define an angle-like variable $w(\chi,u)$, replacing $u$
  and varying it from $0$ to $2\pi m$ on each $\chi$ surface.
\bibitem{FootNote3}
In the finite-cylinder model shunting is unambiguously represented by any
  current streamline that originates from the radial wall and terminates at
  either end of the cylinder.
\end{thebibliography}

\end{document}